\documentclass[Crown,sagev,times]{sagej}

\usepackage{moreverb,url}

\usepackage[colorlinks,bookmarksopen,bookmarksnumbered,citecolor=red,urlcolor=red]{hyperref}

\usepackage{natbib}
\newcommand\BibTeX{{\rmfamily B\kern-.05em \textsc{i\kern-.025em b}\kern-.08em
T\kern-.1667em\lower.7ex\hbox{E}\kern-.125emX}}

\def\volumeyear{2016}
\begin{document}

\runninghead{Marquez and Wakefield}

\title{Harmonizing Child Mortality Data at Disparate Geographic Levels}

\author{Neal Marquez\affilnum{1} and Jon Wakefield\affilnum{2}}

\affiliation{\affilnum{1}Department of Sociology, University of Washington\\
\affilnum{2}Department of Statistics, University of Washington}

\corrauth{Neal Marquez, Department of Sociology
University of Washington,
Seattle, Washington, USA}

\email{nmarquez@uw.edu}

\begin{abstract}
There is an increasing focus on reducing inequalities in health outcomes in developing countries. Subnational variation is of particular interest, with geographically-indexed data being used to understand the spatial risk of detrimental outcomes and to identify who is at greatest risk. While some health surveys provide observations with associated geographic coordinates (point data), many others provide data that have their locations masked and instead only report the strata (polygon information) within which the data resides (masked data). How to harmonize these data sources for spatial analysis has seen previously considered though only ad hoc methods have been previously considered, and comparison of methods is lacking. In this paper, we present a new method for analyzing masked survey data, using a method that is consistent with the data generating process. In addition, we critique two previously proposed approaches to analyzing masked data and illustrate that they are fundamentally flawed methodologically. To validate our method, we compare our approach with previously formulated solutions in several realistic simulation environments in which the underlying structure of the risk field is known. We simulate samples from spatio-temporal fields in a way that mimics the sampling frame implemented in the most common health surveys in low and middle income countries, the Demographic and Health Surveys and Multiple Indicator Cluster Surveys. In simulations, the newly proposed approach outperforms previously proposed approaches in terms of minimizing error while increasing the precision of estimates. The approaches are subsequently compared using child mortality data from the the Dominican Republic where our findings are reinforced. The ability to accurately increase precision of child mortality estimates, and health outcomes in general, by leveraging various types of data, improves our ability to implement precision public health initiatives and better understand the landscape of geographic health inequalities.
\end{abstract}

\keywords{Model-based geostatistics, Bayesian inference, spatial misalignment, spatial demography}

\maketitle

\section{Introduction}
While many have touted the importance of creating more precise approaches for directed health policies in low and middle income countries\citep{Bhutta2016, Desmond-Hellmann2016}, the appropriate approach to assess health variation and hot spots is still not agreed upon. For the majority of low and middle income countries  comprehensive vital registration systems are lacking, and estimating subnational variation of health outcomes from survey sources is the most common strategy for assessing sub-national geographic variation.  Nearly all surveys in this context using two-stage household cluster sampling. Recently there has been an uptick in methods for estimating continuous risk fields for health and demographic outcomes from survey sources \citep{Wakefield2019,Diggle2019}. Estimating continuous risk fields from survey sources is appealing in a number of respects. First, once inference is available for the field, one can aggregate to arbitrary administrative areas, making it an extremely flexible product for policy makers. Second, the start up costs to conduct nationally representative surveys, while large, are much less than for a comprehensive vital registration system and thus methods which can estimate health outcomes at sub-national level from a survey source allow for spatially granular analysis in low resource settings \citep{Alkema2014}.

Estimating continuous spatial risk fields, however, is contingent on having representative survey data which is geolocated. This usually means that data are accompanied by a set of coordinates, most often latitude and longitude. Henceforth, we refer to this type of data as {\it point data}, with the points referring to the cluster locations. While most survey data is representative at some administrative level, by requiring surveys to have point data we further limit the data sources which are available to us for inference on measures of health risks. One such example is the case of child mortality data sources in the Dominican Republic.  Because of a limited vital registration system \citep{Mahapatra2007, Mikkelsen2015}, estimation of child mortality rates within the Dominican Republic has been heavily reliant on the few sub-nationally representative surveys, such as the Demographic and Health Survey (DHS), that have been conducted in the country. While the DHS provides a reliable data set of over 300 surveys constructed with a multi-stage cluster household sampling design in over 90 countries which often include point data, no one country has been surveyed more than five times in the 16 year period between 2000 and 2015, with many countries, like the Dominican Republic, only being surveyed on two or fewer occasions\citep{Burgert-Brucker2016, Gething2015}. This limitation has led to several studies that go beyond the inclusion of solely point data for continuous spatial risk modeling by developing methods which allow for data other than point data to be used in estimating continuous risk fields. 

How best to incorporate non-point data into continuous spatial risk modeling is not agreed upon and to date no study has made a comparative effort to evaluate the different proposed methods. Furthermore, the applicability of methods to different health and demographic data types is not well discussed. Some methods claiming to be applicable to nearly all data types for which areal data can be associated, such as census data \citep{Golding2017}, while others focus on more traditional survey data sources where cluster based sampling is conducted but cluster locations are masked \cite{Utazi2018a}. We pay special attention here to the latter, henceforth referred to as {\it masked data}, which is a common form of data for both the DHS as well as the comparable Multiple Indicator Cluster Survey (MICS).

In this paper, we present a new model for dealing with masked data in the context of estimating a continuous spatial risk surface. The method we develop is consistent with the data-generating process, which is not true of other methods that have been proposed in the literature. We begin our discussion in Section \ref{sec:motivating} with a description  of the data which will be used for the analysis in this paper, both for anchoring our simulation testing framework and also for our substantive application, which is estimating the under-5 mortality rate (U5MR) in the Dominican Republic. In Section \ref{sec:methods} We present the new method for incorporating masked data into a geospatial analysis, along  with a review of other approaches that have attempted to combine point and masked data for this type of analysis and describe why they are flawed. We test the new method in a simulation framework in Section \ref{sec:simulation}, sampling from the spatial field in a number of different scenarios, and comparing the proposed and existing methods in their ability to recreate the underlying field. In Section \ref{sec:results} we apply our method to estimating subnational variation in U5MR for the years2000--2014 in the Dominican Republic using three DHS and MICS surveys. We conclude with a discussion in Section \ref{sec:discussion}.

\section{Motivating Data}\label{sec:motivating}

\begin{figure}[ht]
    \centering
    \includegraphics[width=1.\textwidth]{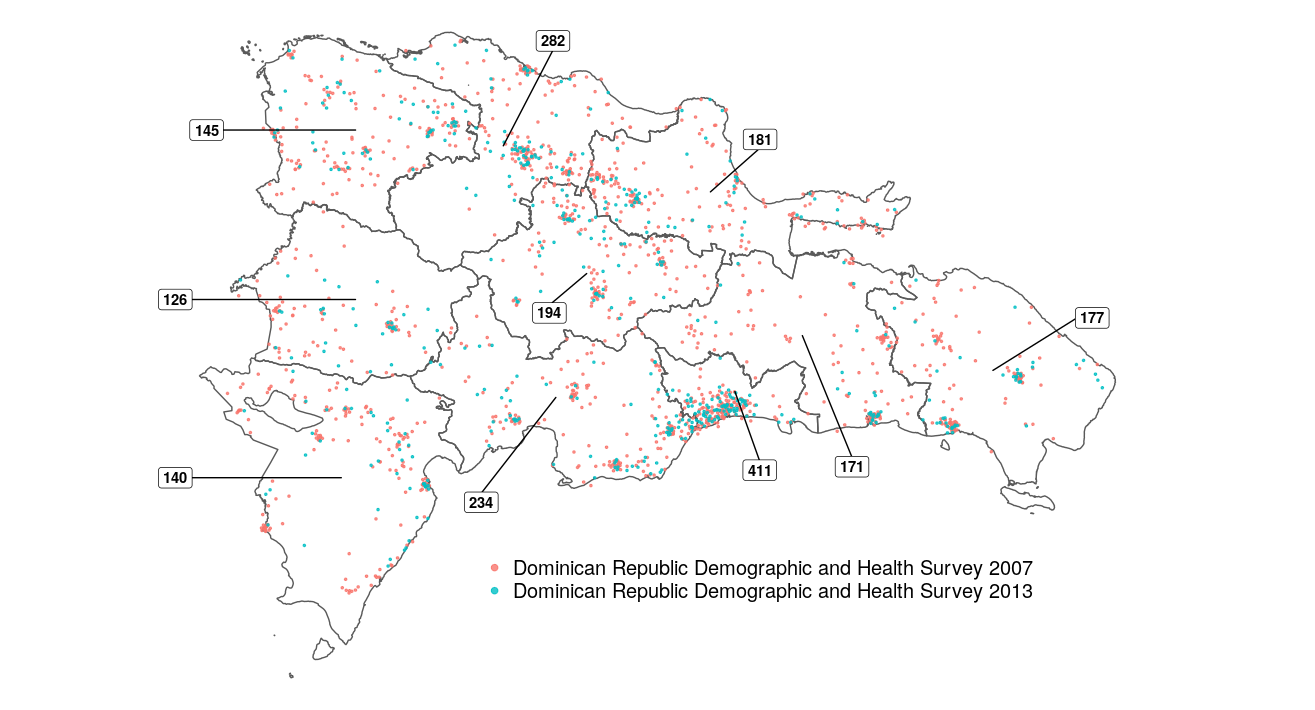}
    \caption{The points represent the (jittered) cluster locations for the 2007 and 2013 DHS. The labeled counts are of the number of masked clusters in each region, 10 regions in total, for the  2014 MICS.}
    \label{fig:samplePlot}
\end{figure}

Data on child mortality in the Dominican Republic are available from two DHS studies conducted in 2007 and 2013 as well as a MICS study conducted in 2014. While the Dominican Republic has had a functioning Vital Registration system for some time, its coverage has not been comprehensive \citep{Mahapatra2007, Mikkelsen2015}. Furthermore, while the census and several in country surveys (including Encuesta Nacional de Hogares de Propositos Multiples 2009) have been conducted, only the DHS and MICS surveys have reliable subnational geographic information on child mortality. Both DHS surveys were collected to be representative at the province level, stratified by urban and rural areas, with a total of 31 provinces and the Distrito Nacional which functions as an independent district. Data was collected using women aged 15 to 49 as the target population. The 2002 and the 2010 census were used to create the sampling frames for the 2007 and 2013 surveys, respectively. From these frames, clusters (enumeration areas) were selected within each of the strata. In each survey, a total of 63 strata were used, one for each of the 31 provinces urban and rural regions and one for the Distrito Nacional which only covers an urban population. Within each strata, clusters were selected for sampling based on the number of households in that strata (using probability proportional to size sampling).  From the potential 35,700 and 39,111 clusters in the country, 1428 and 524 clusters were selected for the 2007 and 2013 DHS, respectively\citep{DHS2007, DHS2013}. GPS coordinates are given for each cluster sampled in the surveys and although these clusters are in fact small areal units we treat them as points in our analysis, as is conventionally done. Urban/rural cluster locations are displaced by up to 2 km/5 km with the locations of a further 1\% random sample of rural clusters being jittered by up to 10 km. Figure \ref{fig:samplePlot} shows the spatial distribution of the clusters at the reported locations. As with other studies we will not consider the jittering further. 

The data extracted from MICS 2014 were collected to be representative of the 15--49 female population for the 10 administrative regions of the Dominican Republic, stratified by urban and rural areas, for a total of 20 strata. Enumeration areas (which we again refer to as clusters) were used from the 2010 census, and 2083 clusters were selected with 33,328 households targeted for surveys\citep{MICS2014}. For the MICS survey, all cluster locations were removed before the data was made publicly available. Only the information regarding the strata from which a cluster resided was provided, as well as which households came from a particular cluster. We emphasize at this point that we are not considering the situation in which the risk (or total count) of a variable is available over a complete area (such as might be available from a census). Methods for this scenario have been considered previously\citep{Wilson2018}. 

Our target of inference is the U5MR, which is the probability of death within the first five years of life. Complete birth histories are collected from women aged 15--49 who reside within households surveyed in both the DHS and MICS. Birth histories include all births for the woman being interviewed, the date of birth of the child, as well as the age in months of a child who died under the age of five. The retrospective nature of the data allow one to reconstruct historical risks since older surveyed women provide information over the complete period in which they gave birth. For this analysis, only children born between 2000 and 2014 were retained for the final dataset. Again following convention, in Section \ref{sec:U5MRmodel} we will describe  a discrete hazards model with a different hazard for 7 distinct age bands. The  7 age periods are: 0 to 1 month (neonatal), 1 to 6 months (post-neonatal period 1), 6 to 12 months (post-neonatal period 2), and 4 annual periods. Each child contributes a number of observations up to all the age groups through which they are  observed. A child does not contribute to all age groups if they die before entering the final age group or if at the time of the survey the child has not reached the final age group. For further details see Wakefield et al.~2019\citep{Wakefield2019}. In our final data set the DHS 2007 and DHS 2013, contain 503,205 and 310,616 person periods, respectively, and the MICS 2014 contains 1,220,781 person periods.

\section{Methods}\label{sec:methods}

\subsection{Mixture Model For Spatially Masked Binary Data}

The data we consider are available as a mix of both point located and masked data. We briefly review the method used for point located data and then introduce our masked data model. For point data, we use a space-time model that has been  previously described\citep{Wakefield2019}. The approach falls under the popular model-based geostatistics approach to estimating a spatial risk surface\citep{Diggle2019}. Under this approach there is an assumed latent continuous spatial field, which is taken as a Gaussian Process (GP), and the parameters of this field are estimated using likelihood or Bayesian methods. There are many ways to carry out computation for a GP model\citep{Heaton2019} and here we follow the stochastic partial differential equations (SPDE) approach\citep{Lindgren2011}, in which the GP is approximated on a triangular mesh to give an efficient Gaussian Markov Random Field (GMRF) representation. 

 We first specify a space-time model for a generic binary indicator for which $p(\boldsymbol{s}_i, t)$ is the probability of an event at a cluster with location $\boldsymbol{s}$ and at time $t$. The spatio-temporal model is:
\begin{eqnarray}
    \text{logit}\left[ ~p(\boldsymbol{s}, t)~\right] &=& \boldsymbol{x}(\boldsymbol{s}, t)^\text{T} \boldsymbol{\beta} + u(\boldsymbol{s}, t)  \nonumber \\
    \boldsymbol{u} &\sim& \mathcal{GP}(\boldsymbol{0}, \mathcal{M} \otimes\text{AR(1)}), \label{eqn:spde} 
\end{eqnarray}
where $\boldsymbol{x}(\boldsymbol{s},t)$ are covariates with associated log odds ratios $\boldsymbol{\beta}$ and $\boldsymbol{u}$ is the  space-time model, which is a GP with a covariance function that is a separable process. Specifically, the covariance function is a combination of a spatial dependence structure, $c_\text{S}$, and a temporal dependence structure, $c_\text{T}$, of the form,
$$
c_{\text{ST}}((\boldsymbol{s}_1,t_1),(\boldsymbol{s}_2,t_2)) = c_\text{S}(\boldsymbol{s}_1,\boldsymbol{s}_2) \times c_\text{T}(t_1,t_2).
$$
The multiplicative structure is beneficial because it is easy to construct a valid spatio-temporal covariance function by combining valid spatial and temporal covariance functions. We choose the spatial component of the separable spatio-temporal model to have Mat{\'e}rn covariance function,
$$
c_\text{S}(\boldsymbol{s}_1,\boldsymbol{s}_2) = \sigma^2 \frac{2^{1-\nu}}{\Gamma (\nu)} \Bigg( \sqrt{8 \nu} \frac{|| \boldsymbol{s}_2 - \boldsymbol{s}_1 ||}{\rho} \Bigg) K_\nu \Bigg( \sqrt{8 \nu} \frac{|| \boldsymbol{s}_2 - \boldsymbol{s}_1 ||}{\rho} \Bigg)
$$
where $\rho$ is the spatial range corresponding to the distance at which the correlation is approximately 0.1, $\sigma$ is the marginal standard deviation, $\nu$ is the smoothness, and $K_\nu$ is a modified Bessel function of the second kind, of order $\nu$. In our model, the Mat\'ern spatial structure is approximated via a SPDE and combined with an autoregressive process of order 1 process in time, denoted AR(1). Inference may be carried out for this model using INLA or TMB. In both approaches, samples can be drawn from an approximation to the posterior distribution, in order to make inference about functions of interest.  Implementation is discussed further in Section \ref{sec:implementation}.
Suppose we observe $Y( \boldsymbol{s},t)$ counts, out of $N( \boldsymbol{s},t)$ samples, with the location of data collection $\boldsymbol{s}$, {\it known}. Then, the above space-time model can be combined with likelihood,
$Y( \boldsymbol{s},t)~| ~p( \boldsymbol{s},t) \sim \mbox{Binomial}[~ N( \boldsymbol{s},t), p( \boldsymbol{s},t)~]$.

Now suppose we have data that has been sampled from $n_k$ clusters with unknown locations, and label such masked data $Y_i(\mathcal{A}_k, t)$ for cluster $i$  in strata $\mathcal{A}_k$, $i=1,\dots,n_k$, for $k=1,\dots,K$ (so that $K$ is the number of strata). We assume there are $m_k$ {\it possible locations} for each of the masked locations.
In this case, under the above binomial sampling model, the implied model for the masked data  is a {\it mixture} model, where we mix (average) over all possible unknown locations, say $\boldsymbol{s}_1,\dots,\boldsymbol{s}_{m_k}$ in strata $ \mathcal{A}_k$ (typically $m_k  \gg n_k$). The likelihood for the masked data labeled $i$ is:
$$
\Pr( Y_i(\mathcal{A}_k, t)~ | ~p(\boldsymbol{s}_1,t),\dots,p(\boldsymbol{s}_{m_k},t) ) =
\sum_{j=1}^{m_k} \Pr(  Y_i(\mathcal{A}_k, t)~ |~ p(\boldsymbol{s}_j,t) ) \times q(\boldsymbol{s}_j, t),
$$
where $$
Y_i(\mathcal{A}_k, t)~ |~ p(\boldsymbol{s}_j,t) \sim \text{Binomial}\left[~N_i,p(\boldsymbol{s}_j,t) ~\right] 
$$
for $i=1,\dots,n_k$ clusters with masked location information in strata $\mathcal{A}_k$, and where $q(\boldsymbol{s}_j, t)$ is the probability of the cluster falling at location $\boldsymbol{s}_j$ at time $t$, $j=1,\dots,m_k$.
To implement this method, the surface that we are estimating should be sufficiently discretized by what we refer to as cells, from which the cluster could have been drawn. Cells are usually constructed in continuous spatial epidemiological modeling to aggregate estimates of risk surfaces to administrative areas. Guidelines for how fine of a discrete surface should be created for this approximation process can be based on the population density data that are available, the number of clusters that are known to exist in the strata (this information is often available in the survey reports) and on the guidelines for mesh creation in the SPDE process\citep{Lindgren2011}. The mixture weights $q(\boldsymbol{s}_j, t)$ can be based on population density, which is available, for example, from WorldPop\citep{Tatem2014, Stevens2015}.
For our simulation studies as well as our analysis of U5MR we will use values of $q$ generated from population rasters made by WorldPop for females age 15--49, which is the age group used for the MICS 2014 dataset.

In an ideal world, we would have the locations of all the clusters in the sampling frame (i.e.,~the enumeration area locations from the census), and we could then choose $q(\boldsymbol{s}_j, t)$ to mimic the actual selection probabilities in the survey. This information will rarely be available, and so we instead, effectively, try to approximate the sampling frame locations using population density at all cells within the strata. This approximation has the added benefit of not needing to adjusted for the changing boundaries of enumeration areas over time. To reiterate, using the population density at the cells  makes sense, since many household surveys use probability proportional to size sampling for the clusters.

\subsection{Previous Approaches}

There have been two previous approaches to incorporating masked data into spatial analysis in a health setting. 
In a recent analysis by Utazi et al\citep{Utazi2018a}, the authors presented a method for use with DHS data in which all locations are masked. 
The method, henceforth referred to as the {\it Ecological} approach, was applied to the estimation of vaccination rates in Afghanistan and Pakistan for a single year where only masked data are available. 
Predictions of vaccination coverage on a grid were produced, based on the masked data only. 
Let $Y_k$ and $N_k$ represent the {\it sums} of the numerators and denominators in strata $k$ (i.e.,~summed over all masked clusters in $k$). 
The key observation here is that these values do not represent the population totals in strata $k$, but rather the sums over the sampled clusters. 
However, the Ecological method that is fitted is in the spirit of observing the totals, and so attempts to obtain the averaged risk. 
This is one key problem with the approach. Unfortunately, a second, and also serious problem with the approach is that the method is susceptible to ecological bias\citep{Wakefield2008} and the change of support problem (COSP)\citep{Gotway2002, Bradley2016}, otherwise known as the modifiable areal unit problem (MAUP)\citep{Wong2009}. 
The Ecological model that was used is,
\begin{eqnarray}\label{eqn:utazi}
    Y_k ~|~ p_k &\sim& \text{Binomial}(N_k , p_k),  \nonumber \\
    \text{logit}(p_k) &=&\left(  |\mathcal{A}_k|^{-1}\int_{\mathcal{A}_k} \boldsymbol{x}(\boldsymbol{s})^\text{T}~d \boldsymbol{s}\right) \boldsymbol{\beta} + \left( |\mathcal{A}_k|^{-1} \int_{\mathcal{A}_k} u(\boldsymbol{s}) ~d \boldsymbol{s}\right)  + \phi_k ,
    \end{eqnarray}

where $\phi_k$ is a discrete spatial random effect (with the Leroux model being used\cite{Leroux2000}), for  $k = 1, \dots K$,  where $|\mathcal{A}_k|$ is the size of strata $k$. The reason that this model is susceptible to ecological bias is that, if the data were a total from the area, then $p_k$ should be evaluated as the average of the risk function, and should not be the risk function evaluated at the averages of the covariates and the average of the spatial field.  We note in passing another, albeit  smaller problem, that the integrals in (\ref{eqn:utazi}) should be weighted by population density. For almost all geographies, populations are not evenly distributed in space and the uneven distribution should be accounted for, as has been done for many continuous spatial modeling post-estimation aggregation endeavors\citep{Wakefield2019, Golding2017, Burstein2019}.
Predictions on a grid $\boldsymbol{s}_g$, $g=1,\dots,n_g$ are obtained via:
\begin{eqnarray}\label{eqn:utazi2}
    \text{logit}(p_g) &=&  \boldsymbol{x}^\text{T}_g \boldsymbol{\beta} + u(\boldsymbol{s}_g) + \phi_{k[g]},
\end{eqnarray}
for $g=1, \dots , n_g$, where $ \boldsymbol{x}^\text{T}_g= \boldsymbol{x}(\boldsymbol{s}_g)^\text{T}$, and $k[g]$ indicates that we pick up the random effect for the strata $k$ that grid point $g$ lies within. In general, when one moves between geographical levels, with a nonlinear model, the regression model changes form, 
i.e.,~if one averages the expit function (at the point level), one does not obtain an expit function. In the present context, this means that the $ \boldsymbol{\beta}$ in (\ref{eqn:utazi}) and (\ref{eqn:utazi2}) are not compatible with each other -- one is an area-level (ecological) association and the other is a point-level association.

The authors state that they are following on from previous work\citep{Moraga2017}, but in the paper quoted, the totals over areas (block averages) were being modeled, and the likelihood was normal with a linear model. For this likelihood, the aggregation is straightforward. The extension to modeling the total over areas for Poisson data has also been previously considered\citep{Wilson2018}. 
We emphasize that this is all in the context of modeling a total, but that is not the situation for the DHS masked data example considered in the paper that introduces the Ecological method\citep{Utazi2018a}.

Another approach which has been applied in several different research applications \citep{Burstein2019} is the method first described by Golding et al.\citep{Golding2017} for child mortality estimation. This method, which we will refer to as the {\it Resample} method, can be thought of as a two-step process, that first restructures the data so that it can be used in a more traditional spatio-temporal modeling approach. The pre-processing takes data that is without a known point location and distributes those points randomly in the given area from which the data was collected. A $k$-means clustering algorithm is used to decide on a set of point locations at which to redistribute the masked data points. Specifically, 10,000 point are randomly assigned to cells within an area relative to the population density in that area and then clustered in space relative to a threshold which can be thought of as the number of people represented at each point. Once the points are $k$-means clustered, the individual observations are assigned to each $k$-means cluster relative to the population that each cluster represents. The threshold for how many $k$-mean clusters are created can be changed depending on the setting, however, most analysis using this approach use a similar threshold for distributing points in an area. 

The Resample method restructures the input data so that it meets the criteria for inclusion into a traditional spatio-temporal model. Unfortunately, however, the method (like the Ecological method) is ad hoc and not based on any model that could be used to simulate masked data. In the Mixture method that we propose, we are averaging over the potential locations of the masked clusters, rather than spreading out the signal from those clusters. The Resample method  ignores the uncertainty of attributing data to a location when that location is unknown. Because data is randomly assigned to a location, model uncertainty is artificially deflated for these locations making estimates look more certain than they actually should be, given that it is not known if data were actually collected from that location. In addition, the relationship between covariates and the outcome could become obfuscated since those variables could also be, and most likely are, wrongly attributed to observations. 

\subsection{Mixture Model for Under Five Mortality in the Dominican Republic}\label{sec:U5MRmodel}

We take as our aim estimating U5MR at the region level over the period 2000--2014. 
For the U5MR simulation and application, we do not have a single binary indicator, but rather a series of binary indicators corresponding to a discrete hazards formulation. We use a modeling approach similar to previous work\citep{Wakefield2019}, though now including data that is location masked and we do not include an additional adjustment for HIV bias correction as prevalence among women in the child bearing age groups was  less than 1.2\% in the Dominican Republic over the time period of interest\citep{Frank2019}. An additional difference is the binning of age groups and the associated parameter effects. We also used the age groups of Burstein et al.\citep{Burstein2019} with random walks across time. 

In addition to the previous Resample method and our Mixture model, we will also fit the model to the point data only -- we label this approach, {\it Ignore}. While we also attempted to fit the Ecological model, we were unable to achieve sensible results and thus removed them from the comparison framework. For each of the Mixture, the Resample and the Ignore approaches we fit four models of various levels of complexity, accounting for different factors that may be important for child mortality. In the model we define, the mortality risk as a function of location $\boldsymbol{s}_i$, time $t$, age $a$, survey $v$, cluster $c$ and the urban/rural status of the cluster. To simplify notation we do not explicitly make the probabilities a function of all of these parameters, but for age $a$ it is important, since the responses recorded are a function of $a$ and so we write $Y_a(\boldsymbol{s}_i, t)$ and $N_a(\boldsymbol{s}_i , t)$ to denote the number of deaths and number of exposure months in age group $a$ at location $\boldsymbol{a}_i$ and in year $t$, with associated hazard (risk conditional on being alive at the start of the age period), $p_a(\boldsymbol{s}_i , t)$.
The models we use are summarized as:
\begin{eqnarray}\label{eqn:u5mmodel}
Y_a(\boldsymbol{s}_i, t) &\sim& \text{Binomial}\left[ ~N_a(\boldsymbol{s}_i , t), p_a(\boldsymbol{s}_i , t)~ \right] \nonumber \\
\text{logit}\left[~ p_a(\boldsymbol{s}_i, t)~\right] &=&
\underbrace{\underbrace{\underbrace{
 \beta_a + 
\gamma I( \boldsymbol{s}_i \in \mbox{ urban })+
\epsilon_v + u(\boldsymbol{s}_i, t)
}_{\text{Model 1}} + \phi_{a, t}
}_{\text{Model 2}} + \eta_i
}_{\text{Model 3}}
\end{eqnarray}
where $I( \boldsymbol{s}_i \in \mbox{ urban })$ is an indicator of whether the cluster lies in an urban region, with $\exp(\gamma)$ the associated odds ratio, for cluster $i$.

Model 1 acts as our baseline model and includes, in addition to the urban/rural effect, a fixed effect $\beta_a$ for each age group, a survey random effect $\epsilon_v \sim_{iid} \mbox{N}(0,\sigma_{\text{v}}^2)$ and a spatial temporal GP effect $u(\boldsymbol{s}_i)$ which is constant over time.  Model 2 accounts, in addition, for possible differences in the trajectories of age specific mortality by adding independent random walks of order 2 for each age group $\phi_{a,t} \sim \mbox{RW2}(\sigma_{\phi}^2)$, with a common variance $\sigma_{\eta}^2$. Model 3 builds on Model 2 and accounts for variation within clusters (excess-binomial variation) through an independent and identically distributed random effect for each cluster $\eta_i\sim_{iid} \mbox{N}(0,\sigma_{\eta}^2) $. In addition we test a model 4, which is similar to model 2 but replaces the spatio-temporal gaussian process with a spataial GP which is constant over time. More details on the model specifications may be found in the Appendix.

\subsection{Implementation}\label{sec:implementation}

 MCMC is notoriously inefficient for Gaussian process models\citep{Filippone2013} and so instead we rely on analytic approximations\citep{Lindgren2011}. The most popular approach to Bayesian modeling for spatial and spatio-temporal data is based on the integrated nested Laplace approximation (INLA)\cite{Rue2009, Blangiardo2015}. We do not go into detail on INLA, as there have been a number of excellent review articles\cite{Rue2017, Bakka2018} and book-length treatments specifically on spatial modeling\cite{Blangiardo2015, Krainski2018}. But briefly, INLA cleverly combines  Laplace approximations for Gaussian random effects and regression coefficients with numerical integration for the remaining parameters (for example, variance-covariance parameters).

The INLA approach is implemented for a class of latent Gaussian models in the {\tt R-INLA} package, with a user-friendly implementation. Our own mixture model does not fall in the required class and so for this model, we use an alternative approach, also based on  Laplace approximations, that is available in the Template Model Builder (TMB) R package \citep{Kristensen2016} and allows user-specified models.  TMB is less familiar to a statistical audience, but has been successfully used in ecological  settings\citep{Thygesen2017, Thorson2015}. TMB does not include the numerical integration step, but rather uses empirical Bayes in which the integrated likelihood is maximized as a function of the hyperparameters. 

In terms of the previously-proposed methods, we generally try to stay as close to possible as the original authors when fitting the models. When running the Ecological method we use the INLA package (as did the proposers of this method) and base our code on the sample coded provided in the original paper\citep{Utazi2018a}. For the Resample method, we worked closely with methodological contributors of the original paper in order to implement the method using TMB. 

Informally, the Laplace approximation relies on the spatial random effects having posteriors that are not too far from being normally distributed, given the hyperparameters. In general, Laplace type methods are not conducive to being applied to mixture models. Therefore, to examine the accuracy for our mixture method, we perform simulations to test whether the Laplace approximation provides similar results to a more robust (albeit computationally expensive) inference method, MCMC. In the simulation scenarios we considered, MCMC and the Laplace approximation provided similar results, particularly for the final estimated field (figures in the appendix).
The time required by the two methods for implementation is striking. While the models estimated using the Laplace approximation took several minutes to run, MCMC models took on the order of days for chains to reach acceptable levels of convergence. The difference was close to a 500 times speed increase when moving from the Laplace approximation to MCMC. All code used to run these analysis may be found on the github page, \url{https://github.com/nmmarquez/PointPolygon}, or is available upon request. 

\section{Simulation Studies}\label{sec:simulation}

To test the performance of our approach and previously proposed methods, we simulate a wide variety of risk fields.
Since we know the true parameters in each simulation, we will be able to measure how well the Mixture, Ecological and Resample methods are able to recover the true risk field.
This is the first comparative evaluation of the performance of models for continuous spatial estimation from point masked data in a known simulation environment. In addition, we also include two additional approaches to calibrate model performance. First, we ignore the masked data to provide one baseline; we refer to this approach as the {\it Ignore} model. Second, and at the other end of the spectrum, we fit a model where the true point location of the masked data is known to the analyst and we may use a  spatial-temporal model on the totality of data. This scenario can only be implemented in a simulation environment where the masking of locations is done only for the sake of testing masked data modeling approaches. This approach will be referred to as {\it Unmasked}, which other methods should strive to approach in terms of accuracy. 
We describe and report the results from two simulation studies, one in the unit-square and one using the geography of the Dominican Republic.

\subsection{Unit-Square Simulation Design}

For our first set of simulations we consider a single spatial field and take the risk of an event at location $\boldsymbol{s}_i$ as
\begin{equation}\label{lin}
\text{logit}[
~p(\boldsymbol{s}_i)~
]=
\beta_0+\beta_1 x(\boldsymbol{s}_i) + u(\boldsymbol{s}_i)
\end{equation}
where $x(\boldsymbol{s}_i)$ is a spatial covariate that will be generated from three different distributions.

We lean heavily on the simulation design described by Utazi et al.\citep{Utazi2018a}. To generate risk fields, we begin with a 60$\times$60 cell grid. We assume that the risk for each cell in the grid is generated by an inverse logit of an intercept, one covariate, and a latent spatial field. We create variety in the fields, by varying the following: covariate type, covariate coefficient value, and spatial range. The intercept we hold constant at $\beta_0=-2$, for all fields. Since in practice, a variety of covariates generate real-life risk fields, we consider three types: an independent and identically distributed covariate, a covariate correlated across space, and a covariate constant across large geographic regions. Examples of these fields may be found in the Appendix.
For each of these covariate types, we generate fields with $\beta_1=2$. We vary the spatial range (the distance at which correlation of the latent field drops to 0.1), over the values $\{0.3, 0.5, 0.7\}$. 

We create artificial areal units at which to carry out masking -- we label these square units, the {\it sampling grids}.
These sampling grids divide the spatial field into smaller sub-units that are analogous to areas in the DHS and MICS sampling context. The sampling grids are either 3$\times$3, 5$\times$5, or 10$\times$10 generating 9, 25 and 100 sub-units respectively. We also consider two different {\it sampling types} which describe  whether data is masked or not. In the {\it overlap} scenario, data that is masked and data with point known locations are simulated from the same sub-units while in the {\it non-overlapping} scenario, sub-units have data points that are either point known or point masked but not both.
We sample from each of these generated risk fields to create 54 datasets for model testing  (3 types of covariate, 3 spatial correlations, 3 sampling grids, 2 sampling types). 
We sample from the field in such a way that a total of 300 clusters are observed, with half the clusters coming from known locations and the other half being masked. At each cluster we carry out binomial sampling with denominators of size 250.  For each of the 54 simulation scenarios a field is generated and then we carry out binomial sampling at the sampling clusters 250 times, and the results we present are averaged over these 54.

After creating this simulation environment, we fit our five candidate approaches on each of the 54 datasets. Model performance is assessed over a 60$\times$60 grid and with respect to the  known probability field using a number of validity metrics. Specifically, the root mean squared error (RMSE), coverage of 95\% credible/confidence interval, and the bias of the risk estimates are calculated over the grid to compare models. Let $p(\boldsymbol{s})^{(k)}$ be the posterior median of the risk at location $\boldsymbol{s}$ in simulation $k$, $k=1,\dots,250$. The RMSE is
$$\mbox{RMSE} = \left[
\frac{1}{250} \sum_{k=1}^{250} \left( \hat p(\boldsymbol{s})^{(k)} - p(\boldsymbol{s}) \right)^2
\right]^{1/2},$$
and the bias is,
$$\mbox{BIAS} = 
\frac{1}{250} \sum_{k=1}^{250} \hat p(\boldsymbol{s})^{(k)} - p(\boldsymbol{s}).$$
For presentation, we divide our results by covariate type and spatial range.


\subsection{Unit-Square Simulation Results}

\begin{figure}[ht]
    \centering
    \includegraphics[width=1.0\textwidth]{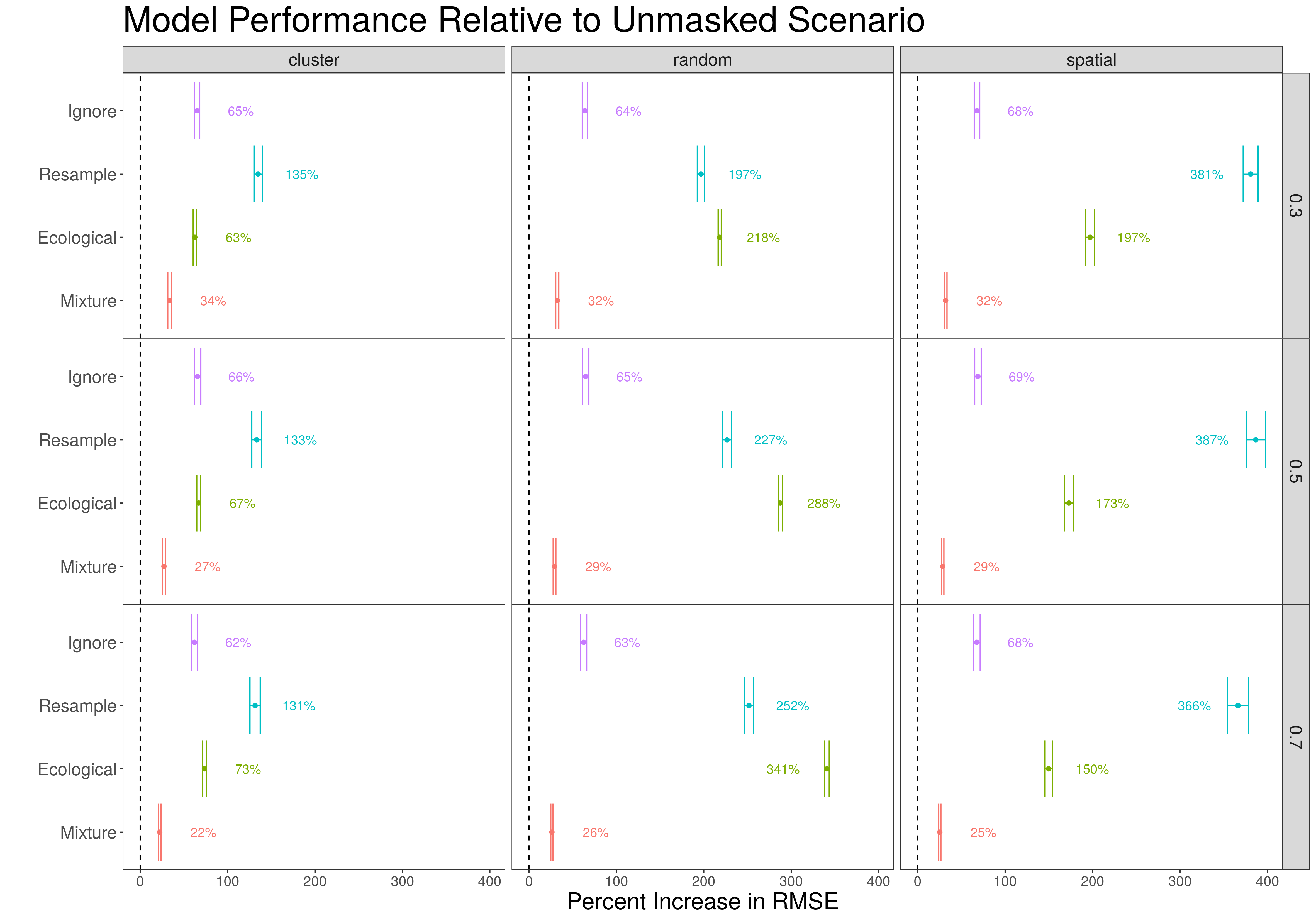}
    \caption{Model performance relative to the gold standard of an unmasked method. Values represent the percent increase of the RMSE compared to the unmasked method. A value of 0\% indicates a method which has the same RMSE as the unmasked method while 100\% indicates a method whose RMSE is twice that of the unmasked method. The vertical bars represent the 2.5 and 97.5 percentiles of results.}
    \label{fig:gridResults}
\end{figure}

In all testing environments for the unit square simulations, we find that our proposed Mixture method outperforms previously presented methods in terms of each of the evaluation diagnostics and is a significant improvement over ignoring the spatially masked data. Results in Figure \ref{fig:gridResults} show the percent increase in RMSE for each of the methods as compared to the Unmasked method which should be seen as the gold standard. The percent increase in RMSE is aggregated over all simulation scenarios for a given covariate type (across columns), and spatial correlation (across rows). The vertical bars show the  lower 2.5th and 97.5th quantiles of the percent increase, across simulations. Note that a value of 0\% indicates that a method performed just as well as the unmasked method,  while a value of 100 \% indicates a method whose average RMSE for that covariate type spatial extent combination is twice as high as the Unmasked method. Across all simulations, the mixture model never exceeds a value of 35\%.
 
Surprisingly, the next best performing method in all covariate type/spatial correlation combinations but one was the Ignore method. Despite the Resample and Ecological methods leveraging more data than the ignore method, they consistently perform worse than simply ignoring data that is not point located.  For the bias metric, the Ecological methods is the poorest and the Mixture model performs best with the Resample method a little behind. In terms  of coverage, the Mixture and Ignore methods perform nearly identically (see Appendix), and are closest to 95\%, as well as the smallest variation around the true 95\% coverage, when compared to the candidate models (apart from the Unmasked case). For both the Resample and the Mixture methods, the estimated 95\% estimation intervals of the field rarely actually covered 95\% of the field.

\subsection{Dominican Republic Simulation Design}

In addition to running simulations on a controlled spatial and sampling environment, we are also concerned with how performance differs in a environment where the sampling frame is fixed for an outcome. In the case of sampling child mortality data in the Dominican Republic, we simulate a spatial risk field over the country and sample in a manner that mimics the sampling frame of the MICS and DHS surveys that took place in the country. By doing so we are able to see which model is able to best reconstruct simulated fields, conditional on the spatial distribution of the population and the locations of samples either up to a given geolocation (point data) as specified in the DHS or to a given stratified administrative areal unit such as in the MICS (masked data). Our simulated field across the Dominican Republic contains 4532 cells where each cell is approximately a 5$\times$5 kilometer unit. Because birth history data also provides historical information on child mortality we simulate a spatio-temporal process over the period 2000--2015. Death can occur in one of 6 intervals, namely, the first month of life, months 2--6, 7--12, 13--24, 25--36, 37--48, 49--60. In our simulation we have a combination of point  and masked data for the first five years of the simulation while the last year has only masked data available for analysis, mimicking the data structure of the 2013 DHS and MICS 2014 modules. 

\begin{figure}[ht]
    \centering
    \includegraphics[width=1.0\textwidth]{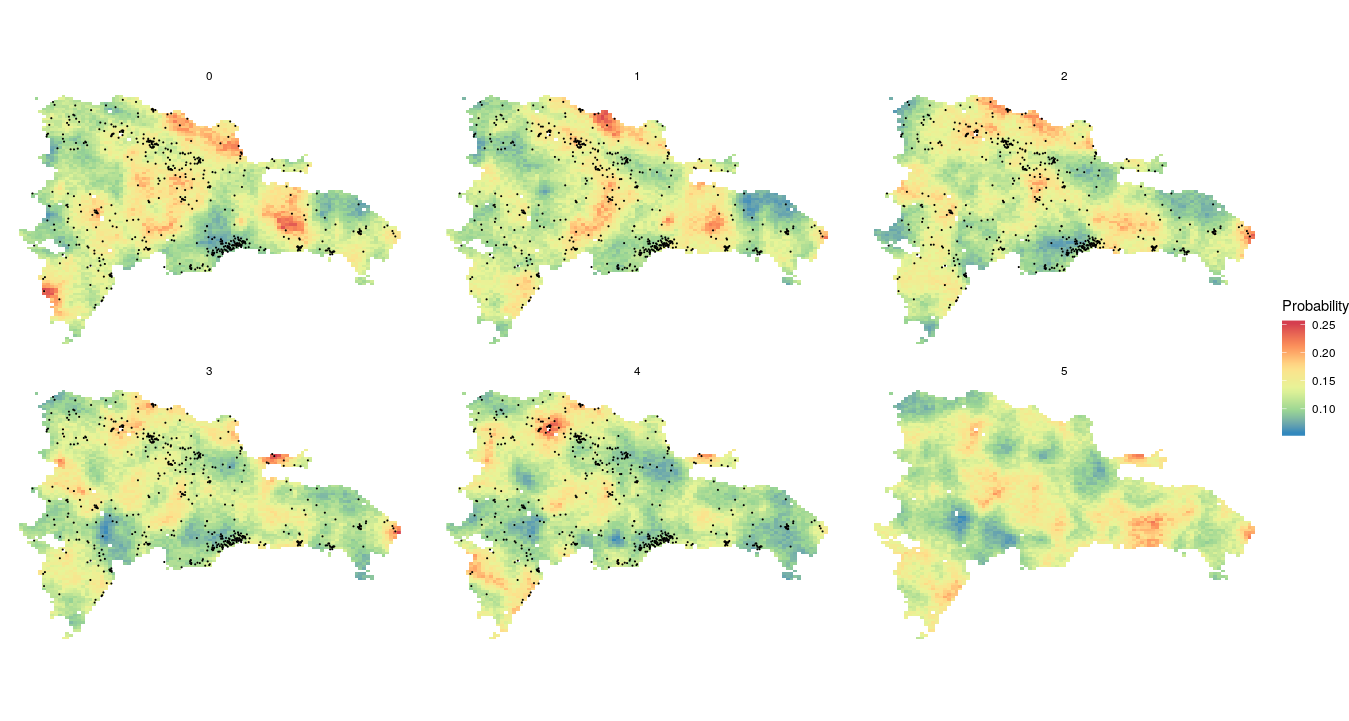}
    \caption{One example simulation risk field on the Dominican Republic over time, along with cluster sample locations, which are directly pulled from the 2013 DHS.}
    \label{fig:drSim}
\end{figure}

We plot the geolocations of clusters that were sampled across the Dominican Republic from the 2013 DHS survey as an example of a simulation field in Figure \ref{fig:drSim}. 
Complete birth histories give us retrospective data for each mother interviewed; we generate our data to have geolocated point observations for years from 2000 to 2013. As stated, MICS data is masked so one only knows the geographical strata in which the sampled cluster lies and not the true cluster location. For each simulation in this experiment we simulate true cluster locations for clusters from the MICS survey, and generate samples at the locations $\boldsymbol{s}_i$ and time points $t$. Cells with a greater population of women ages 15--60 have a higher probability of being selected for cluster placement. 
The probability that a cluster is placed in any given 5$\times$5 kilometer cell is $N_{\text{cell},g}/N_{\text{strata},g}$ where $N_{\text{cell},g}$ is the population of women age 15--60 in cell $g$ (based on the WorldPop estimate) and $N_{\text{strata},g}$ is the population of women age 15--60 in the strata within which cell $g$ is located. Cells may contain more than one cluster, as clusters need not be more than 5 kilometers apart. 

Because the surveys used stratified sampling by region and urban/rural, we extend the method described by Utazi et al\citep{Utazi2018a}. In particular,  the Ecological  method  we use is a conditional auto-regressive (CAR) model  with an adjacency matrix for the regions represented in the MICS survey,  but including a fixed effect for urban/rural status.

The underlying spatio-temporal field is simulated using a formulation similar to the previous grid simulation and equation (\ref{lin}). The difference is that the underlying latent field changes over space and time (over 6 hypothetical years of data) with a precision matrix that is the Kronecker product of an AR(1) time series model and a Mat\'ern spatial GP,  as described in equation (\ref{eqn:spde}). This model is similar to the model we use to estimate U5MR. The variance for the process is fixed at 1 and the temporal AR(1) correlation is set to 0.95 for all simulations of the spatio-temporal field. Random values for the covariate effects, covariate types, spatial range of the geographic correlation process, and the placement of the clusters in our simulation, are altered for each simulation in the same manor as the grid simulation. We do not, however, change the sample size, grid, or types, as done in the previous simulation, as these values are dependent on the clusters sampled in the DHS and MICS modules. In total we create 45 risk fields, and repeat this process 10 times for additional samples -- we only perform 10 simulations as these models are very computationally expensive to run. We compare models using the same performance metrics as for the unit square simulation at the cell level. In addition, we also aggregate risk fields to the 31 province level averages and compare how population weighted aggregated estimates compare with true known values. Typically, small area aggregate measures are more useful for evaluating progress and policy making decisions\citep{Wakefield2019,Burstein2019} and thus comparing how models fair in their ability to recover the true value of health risks at a administrative level is of great importance. We use the same metrics as for the cell level validation in order to assess the performance at the regional level.

\subsection{Dominican Republic Simulation Results}

We find similar results for the conditional simulation scenario compared to the unit-square simulation. The Mixture method outperforms the Resample method, the Ecological  method, and ignoring the point masked data, for both the coverage and RMSE metrics. Interestingly, the Resample method appears to be somewhat improved over the Ecological method, in contrast to the first simulation.

\begin{figure}[ht]
    \centering
    \includegraphics[width=1.0\textwidth]{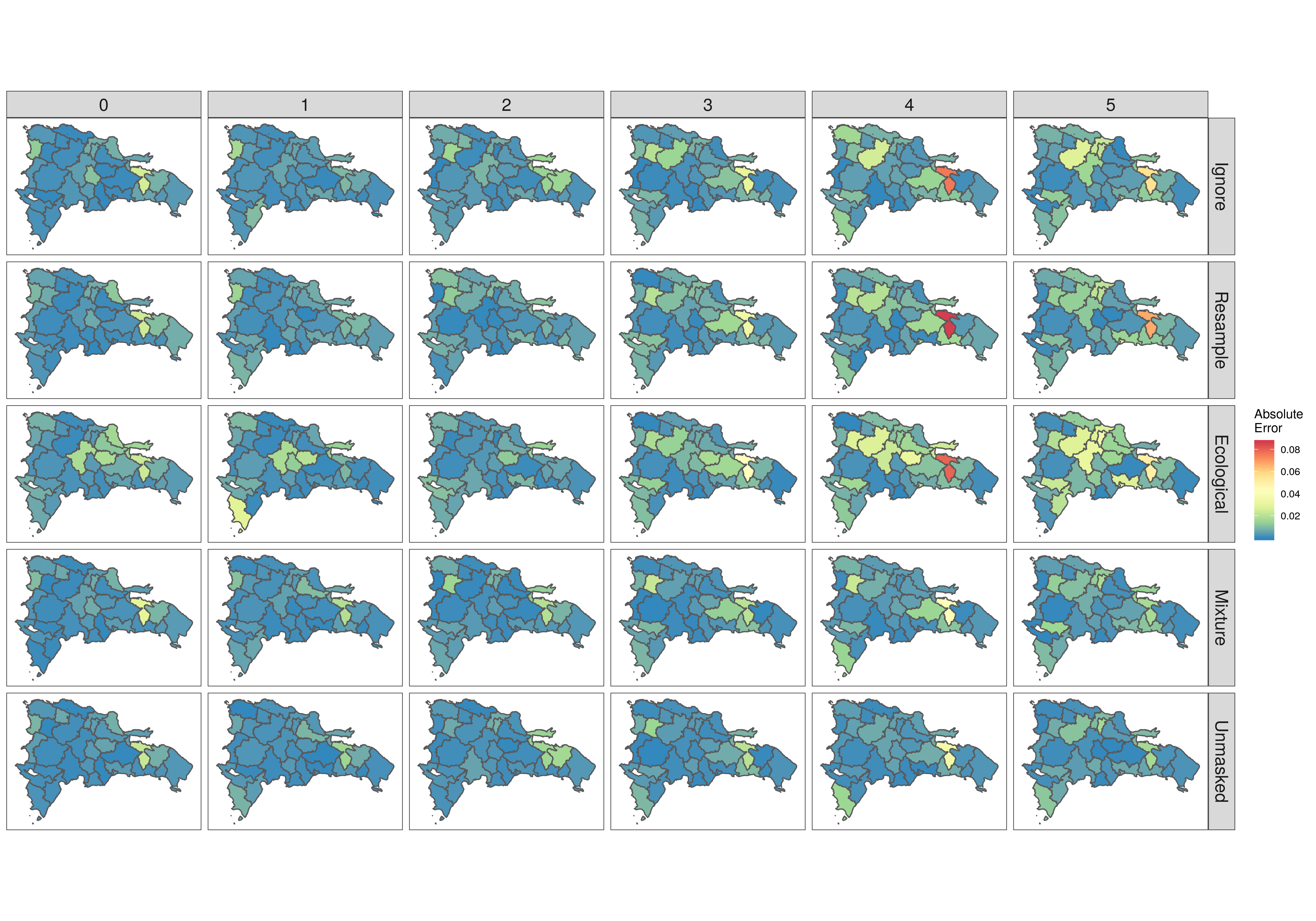}
    \caption{Comparison of Absolute Error at the Province level for Dominican Republic simulation. Columns represent 6 different years of analysis and predictions while rows represent the different methodological approaches. Higher values indicate worse absolute error.}
    \label{fig:drProvMap}
\end{figure}

Model improvement over the Ecological  method ranged from 7\% to 16\% average improvement over the combinations of spatial range and covariate type (figures for cell level comparison may be found in the Appendix). When aggregated to the province level we find that discrepancies between the Ecological method and the Mixture method are exacerbated. We find that the RMSE of province level risk is on average more improved upon than at the cell level when comparing the Ecological method to the Mixture method, with results improving up to 25\% on average under different combinations of spatial range and covariate type. Comparisons at the province level of RMSE for a single simulation across all time periods are shown in Figure \ref{fig:drProvMap}, with the columns corresponding to years of data, labeled 0 through 5. Coverage estimates at both the cell level and the province level are more reliable and unbiased for the Mixture method, compared to both the Resample method and ignoring the masked polygon data. Comparisons of bias of estimates and bias of dissimilarity measures at both province and the cell level may be found in the Appendix. The bias results show that the Resample method has a larger range of bias but is comparable in terms of the average bias to the Mixture model, while the Ecological model shows poor average bias with a large spread across simulations.

\section{Motivating Data Revisited: Results for Under-Five Mortality in the Dominican Republic}\label{sec:results}

We now turn to the motivating example of estimating yearly U5MR in the Dominican Republic. We fit Models 1--3 as given by equation (\ref{eqn:u5mmodel}) and we add a Model 4 which corresponds to Model 2, with the $u(\boldsymbol{s},t)$ replaced by $u(\boldsymbol{s})$, i.e.,~a space only GP. For each of these models we use the Resample and Mixture approaches to deal with the masked data. We also fit Models 1--4 with the Ignore approach that leaves the masked data out.
For each of 12 these candidate fits, we assess performance by fitting the model to all data except for a single year of data (2000--2012) that is held out from model fitting (the training data) and only used for model validation. We do not run a validation test for the year 2013 as the number of observations in that year is relatively small nor 2014 as their is only masked data for this year. We evaluate the performance of a model by calculating the log conditional predictive ordinate (CPO), which corresponds to $ \sum_{t=1}^T \log p(\boldsymbol{y}_t | \boldsymbol{y}_{-t})$, for the test data, for each approach. Full details are given in the Appendix. We leave out all data for a year (including masked data), but only predict for the left-out point data. Overall the temporal mixture model is the most consistent best performing model, accounting for the best performance of all possible model fitting strategies for 6 of the 13 hold out years. Collectively, Resample models were the best performing model for 5 years but overall, averaging overall all years, the Resample approach was poorest and the Mixture approach was clearly the model with the best performance, with the Ignore approach the next most favorable.


 \begin{figure}[ht]
    \centering
    \includegraphics[width=1.0\textwidth]{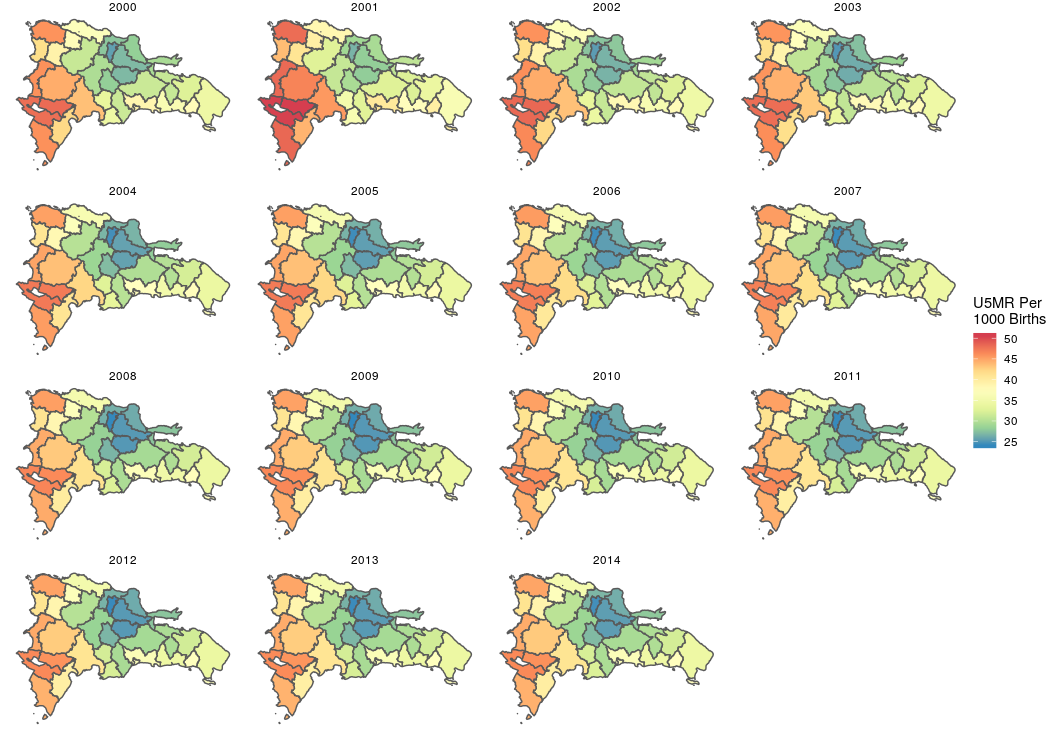}
    \caption{Province level estimates of U5MR per 1000 births, for the Mixture Model.}
    \label{fig:finalMap}
\end{figure}

Figure \ref{fig:finalMap} shows province level U5MR estimates from 2000 to 2014 for the Temporal Mixture model. The Dominican Republic has seen dramatic changes in U5MR since the early 1960's however more recently changes are less striking across the country, as reported by a number of sources\citep{Wang2012, Dicker2018}. Our estimates are consistent with previous findings, as there is not strong temporal variation in the estimates. However, there is persistent strong geographic variation. The west of the Dominican Republic has higher levels of U5MR and the central north has the lowest levels. If we compare the results to the Resample model, we see that there is a different temporal result, with a significant decline in U5MR since 2000, despite having similar spatial variation in mortality. Our estimates, which show a smaller drop over 2000--2015, are more consistent with other organizations efforts to model under 5 mortality at the national level, as we now discuss.

\begin{figure}[ht]
    \centering
    \includegraphics[width=.9\textwidth]{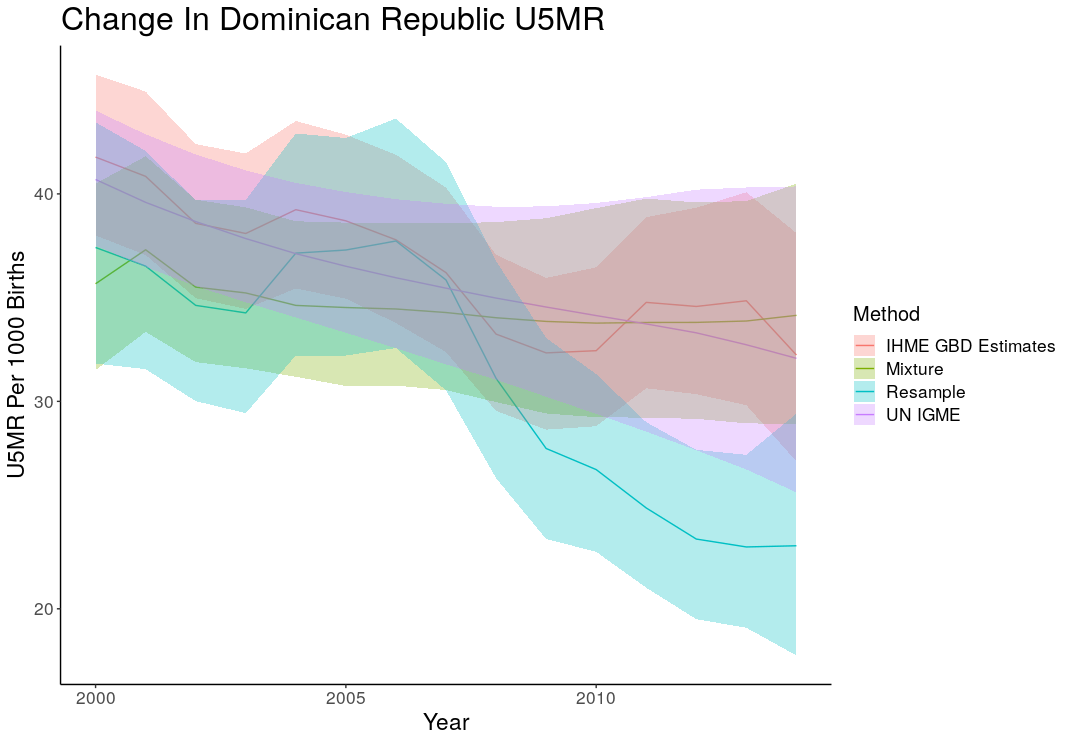}
    \caption{National level estimates of the U5MR over time, from different models.}
    \label{fig:finalTrend}
\end{figure}

In order to get a better point of comparison we show results for both models, aggregated to the national level, in Figure \ref{fig:finalTrend}, along with other studies and publications which estimate child mortality within the Dominican Republic at this level. We include for comparison the Global Burden of Disease 2017 national level results\citep{Dicker2018}. In the appendix we include as well comparisons to the direct estimates from the 2013 DHS and 2014 MICS using Horvitz Thompson (HT) weighted estimators\citep{Horvitz1952} used in other spatial child mortality contexts\citep{Mercer2015} as well as smoothed national level estimates\citep{Mercer2015}. We see that all models, except for the Resample model a much shallower temporal trend in the U5MR between 2000 and 2014. The potential reasons for these differences are many-fold. However, the Resample method can not be statistically justified from a methodological viewpoint and it had inferior performance when compare to the the Mixture model in both of the simulation scenarios. Further, it was also inferior in the sample validation tests. So we do not believe it advisable to accept the Resample results, which differ from every other model of child mortality that we have presented. 

\section{Discussion}\label{sec:discussion}

Precise estimates of epidemiological and public health outcomes are vital for targeted policy making to improve the standing of those that lie at the margins of good health. In an attempt to provide better fine scale geographic estimates of measures such as the child mortality rate\citep{Golding2017,Burstein2019} and vaccination coverage\citep{Utazi2018a}, researchers have proposed methods to incorporate not only point located data but also masked data. To date, however, these methods have not been compared for accuracy and validity, and neither the Ecological nor the Resample methods accurately reflect any feasible data generating process.

In this paper we present an alternative method for incorporating polygon data into continuous spatial analysis. We build two spatial simulation scenarios, one generalized scenario simulating continuous spatial fields on an artificial 1$\times$1 study area that is common for simulations in the spatial literature and another reflecting a typical scenario in global health and epidemiology research which use DHS and MICS surveys. For each simulation scenario we build various spatial fields and compare our new Mixture method to the Ecological and Resample methods. We found in all simulations, via a number of validation techniques, that our newly presented Mixture method outperformed both the other methods and approaches which neglect to incorporate the masked data and use point located data only.

In addition, we also fit a variety  of increasingly complex versions of the new Mixture method, the Resample method, and a traditional continuous spatial field model, to child mortality data collected from MICS and DHS surveys. We perform out of sample validation examinations by withholding one year of data from the model fitting process and assessing a model based on how well it was able to predict the out of sample data. For a majority of annual holdout tests we find that variations of the Mixture method outperform other methods tested in terms of out of sample predictive validity. Furthermore, when using the entire data set, the Mixture method produces results at the national level that are more consistent with previous estimates of the U5MR in the Dominican Republic than are estimates from the Resample method.

In the application, where we do find similarities in all models is in the spatial variation that exists in any given year. While the degree of these differences vary from model to model, all results show especially pronounced differences between eastern and western provinces. This finding is corroborated by the results in  Burstein et al\citep{Burstein2019}. Though their method of spatial analysis follows the Resample paradigm, their final results are adjusted such that they are consistent with the Global Burden of Disease 2017 Dominican Republic child mortality results, and thus show little change over time in the country level estimates of child mortality.

Ensuring that we are accurately incorporating data into our modeling process is vital to producing valid estimates of public health phenomena, especially when the results are to be used for targeted policy making. In our analysis we show that our newly presented Mixture method is able to better leverage data that are not typically suited for continuous spatial analysis, to improve the accuracy of our results. We acknowledge that the scenario considered only represents one specific type of areal data, data that was originally collected at the cluster level, but with the locations masked from the end user. Many other types of polygon data do exist, such as average risk over administrative data found from vital registration records or censuses. This situation has been considered previously\citep{Wilson2018} but not for binomial models. We also do not consider the effect of jittering. Nonetheless, the type of masked data presented in this paper represents a substantial portion of traditionally used data sources for health and demographic outcomes, such as the DHS and MICS.

\section*{Acknowledgments}

Wakefield was supported by National Institutes of Health grants R01CA095994  and R01AI029168. Nathaniel Henry, Aaron Osgood-Zimmerman, and Katie Wilson have provided valuable feedback during the course of this research.


\newpage
\appendix
\section{Appendix}
\subsection{Appendix A: Model Specifications}
\subsubsection{Grid Square Simulation}

\begin{align}
\Pr( Y_i(\mathcal{A}_k)~ | ~p(\boldsymbol{s}_1),\dots,p(\boldsymbol{s}_{M_k}) ) &=
\sum_{j=1}^{m_i} \Pr(  Y_i(\mathcal{A}_k)~ |~ p(\boldsymbol{s}_j) ) \times q(\boldsymbol{s}_j) \nonumber \\
Y_i (s_i) &\sim \text{Binomial} \left[ N_i, p(\boldsymbol{s}_i)\right] \nonumber \\
\text{logit}\left[ ~p(\boldsymbol{s}_i)~\right] &= \boldsymbol{x}(\boldsymbol{s_i})^\text{T} \boldsymbol{\beta} + u(\boldsymbol{s}_i) \nonumber \\
\boldsymbol{u} &\sim \mathcal{GP}(\boldsymbol{0}, \mathcal{M}) \nonumber \\
\mathcal{M} &= \mathcal{M}(\tau, \kappa) \nonumber \\
\beta &\sim \mathcal{N}(0, 100) \nonumber \\
\tau &\sim \text{Log Normal}(0, 100) \nonumber \\
\kappa &\sim \text{Log Normal}(0, 100) \nonumber 
\end{align}

\subsubsection{Dominican Republic Simulation}

\begin{align}
\Pr( Y_i(\mathcal{A}_k, t)~ | ~p(\boldsymbol{s}_1, t),\dots,p(\boldsymbol{s}_{M_k}, t) ) &=
\sum_{j=1}^{m_i} \Pr(  Y_i(\mathcal{A}_k, t)~ |~ p(\boldsymbol{s}_j, t) ) \times q(\boldsymbol{s}_j, t) \nonumber \\
Y_i (s_i, t) &\sim \text{Binomial} \left[ N_i, p(\boldsymbol{s}_i, t)\right] \nonumber \\
\text{logit}\left[ ~p(\boldsymbol{s}_i, t)~\right] &= \boldsymbol{x}(\boldsymbol{s_i}, t)^\text{T} \boldsymbol{\beta} + u(\boldsymbol{s}_i, t) \nonumber \\
\boldsymbol{u} &\sim \mathcal{GP}(\boldsymbol{0}, \mathcal{M} \otimes \text{AR(1)}) \nonumber \\
\mathcal{M} &= \mathcal{M}(\tau, \kappa) \nonumber\\
\text{AR(1)} &= \text{AR(1)}(\zeta) \nonumber\\
\beta &\sim \mathcal{N}(0, 100) \nonumber \\
\tau &\sim \text{Log Normal}(0, 100) \nonumber \\
\kappa &\sim \text{Log Normal}(0, 100) \nonumber \\
\zeta &\sim \text{Logit Normal}(0, 100) \nonumber
\end{align}

\subsubsection{Dominican Republic Child Mortality Estimation}

\begin{align}
\Pr( Y_i(\mathcal{A}_k, t)~ | ~p(\boldsymbol{s}_1, t),\dots,p(\boldsymbol{s}_{M_k}, t) ) &=
\sum_{j=1}^{m_i} \Pr(  Y_i(\mathcal{A}_k, t)~ |~ p(\boldsymbol{s}_j, t) ) \times q(\boldsymbol{s}_j, t) \nonumber \\
Y_i (s_i, t) &\sim \text{Binomial} \left[ N_i, p(\boldsymbol{s}_i, t)\right] \nonumber \\
\text{logit}\left[ ~p(\boldsymbol{s}_i, t)~\right] &= \beta_a + 
\gamma I( \boldsymbol{s}_i \in \mbox{ urban })+
\epsilon_v + u(\boldsymbol{s}_i, t) + \phi_{a, t} + \eta_c \nonumber \\
\boldsymbol{u} &\sim \mathcal{GP}(\boldsymbol{0}, \mathcal{M} \otimes \text{AR(1)}) \nonumber \\
\mathcal{M} &= \mathcal{M}(\tau, \kappa) \nonumber\\
\text{AR(1)} &= \text{AR(1)}(\zeta) \nonumber\\
\beta &\sim \mathcal{N}(0, 100) \nonumber \\
\epsilon_v &\sim \mathcal{N}(0, \sigma_\epsilon) \nonumber \\
\boldsymbol{\phi}_a &\sim \text{RW1}(0, \sigma_\phi) \\
\nu_c &\sim \mathcal{N}(0, \sigma_\nu) \nonumber \\
\sigma_\epsilon &\sim \text{Log Normal}(0, 100) \nonumber \\
\sigma_\phi &\sim \text{Log Normal}(0, 100) \nonumber \\
\sigma_\nu &\sim \text{Log Normal}(0, 100) \nonumber \\
\tau &\sim \text{Log Normal}(0, 100) \nonumber \\
\kappa &\sim \text{Log Normal}(0, 100) \nonumber \\
\zeta &\sim \text{Logit Normal}(0, 100) \nonumber
\end{align}

\subsection{Appendix B: MCMC and Laplace Approximation}

To assess the validity of using a Lpalace Approximation for the mixture model estimation process we took a single dataset from the unit square simulation scenario and estimated the model using both a Laplace approximation and using MCMC. We use the same model fitting process for the Laplace Approximation explained previously. For the MCMC portion we fit the model using STAN and a no U-turn sampler. Four chains were fit and the mean posterior of the field  probability estimates for each chain were compared to the mean estimated probability of the fitted Laplace Approximation fit. In addition we visually compared the posterior standard deviation of the probability field to that of the Laplace fit and found all estimates to be similar. Despite the models producing similar results the run times are drastically different. While the Laplace Approximation took less than 2 minutes to fit each MCMC chain took 4 days to run 2000 steps and sensible convergence was reached. Fitting models using MCMC is infeasible in our entire simulation framework where we are comparing thousands of fit models, however, with the nearly 4000 times speed up from the Laplace approximation we are able to test a wider array of simulation scenarios to assess model performance.   

\newpage

\begin{figure}[ht]
    \centering
    \includegraphics[width=1.0\textwidth]{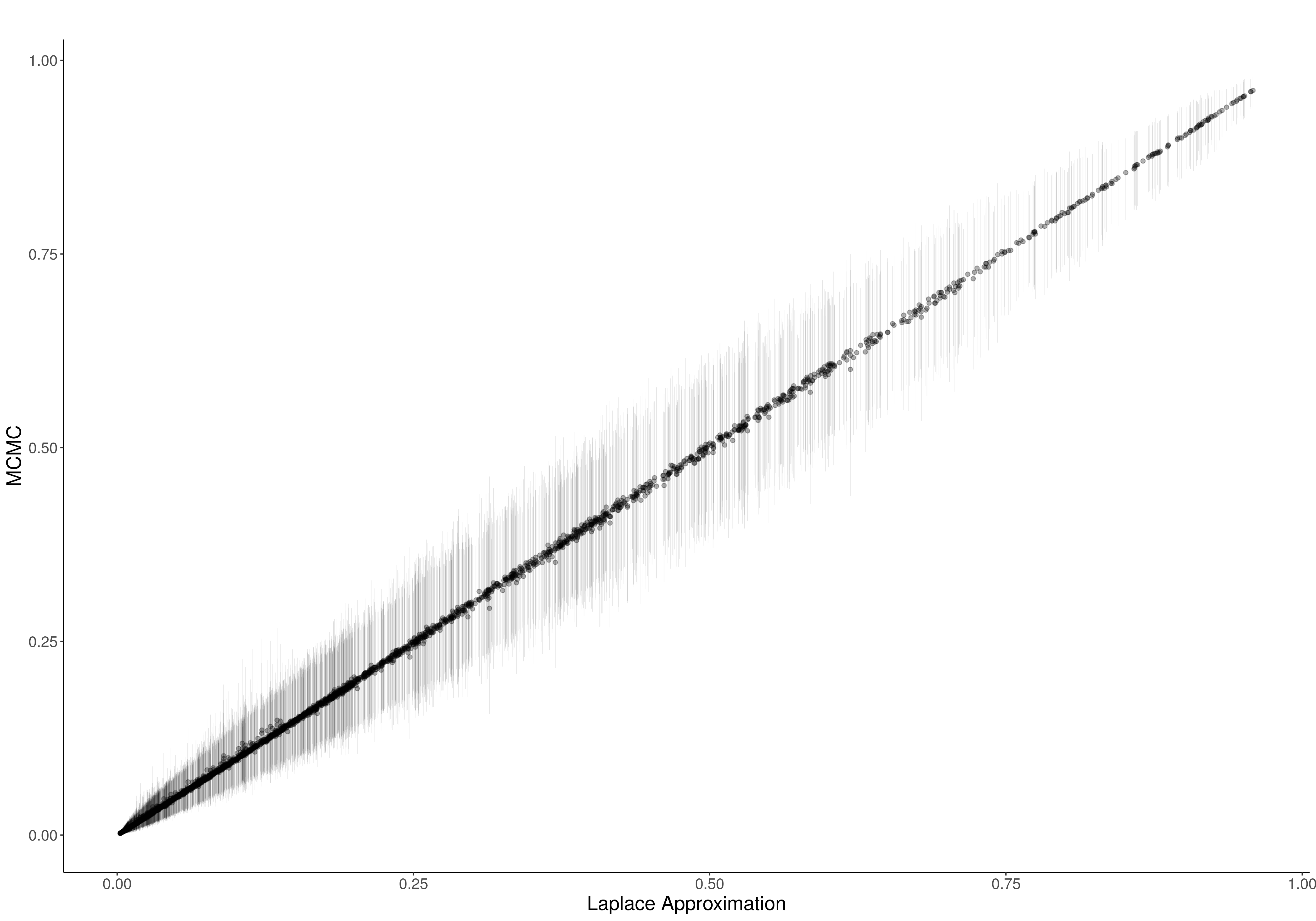}
    \caption{Mean and 95\% credible interval for posterior of estimated probability field for one MCMC chain compared to mean estimate of Laplace Approximation fit.}
    \label{fig:mcmc1}
\end{figure}

\newpage

\begin{figure}[ht]
    \centering
    \includegraphics[width=1.0\textwidth]{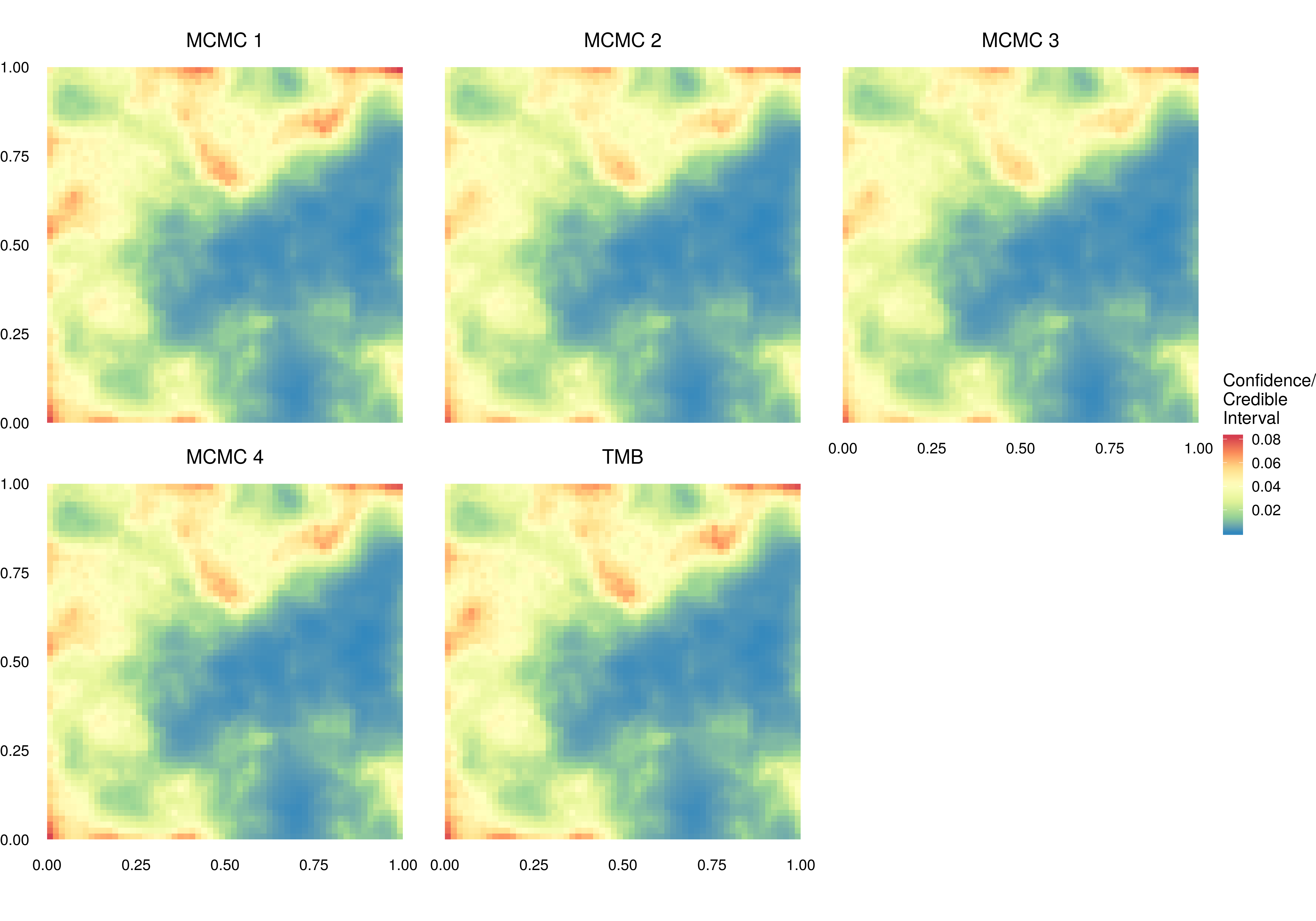}
    \caption{Standard deviation of posterior and standard error of probability for MCMC and Laplace Approximation using Template Model Builder respectively.}
    \label{fig:mcmc2}
\end{figure}

\newpage

\subsection{Appendix C: Unit-Grid Square Simulation Results}

\begin{figure}[ht]
    \centering
    \includegraphics[width=.9\textwidth]{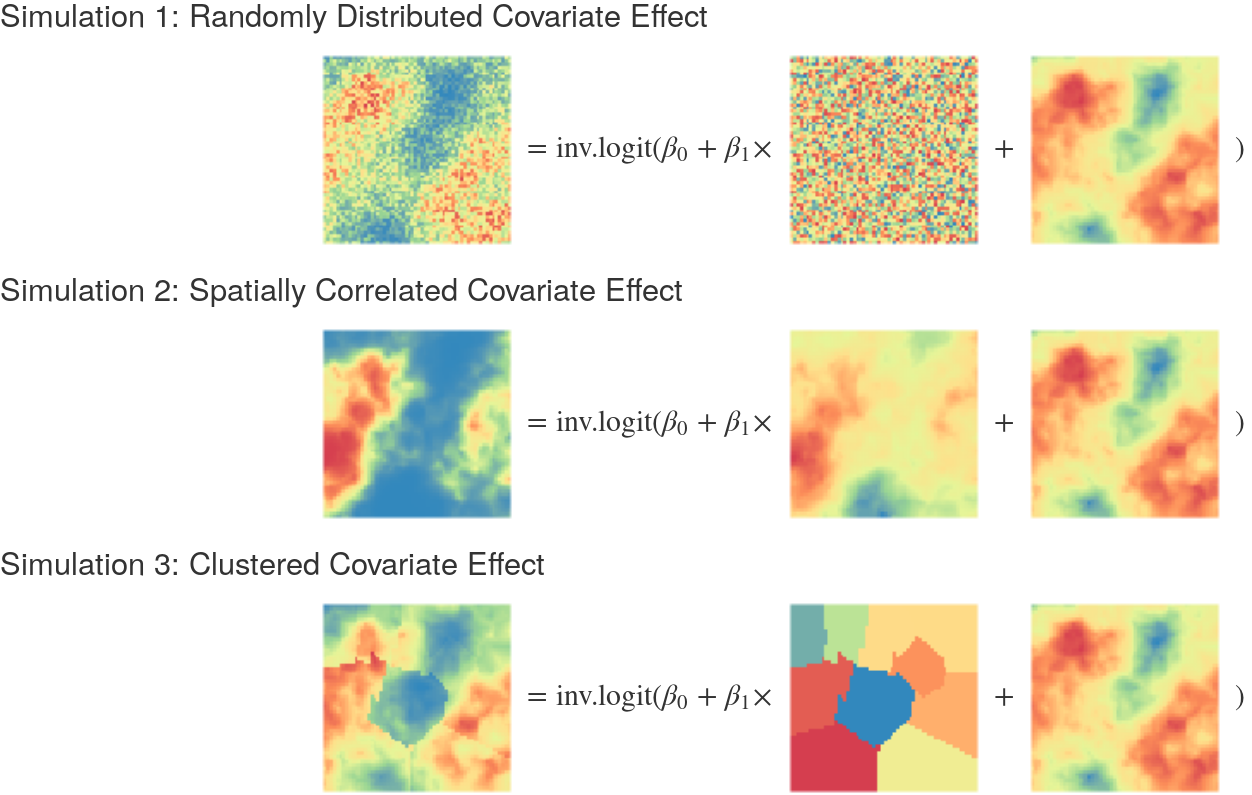}
    \caption{Examples of three simulated risk fields using each of the three covariate types and the same latent spatial field. The risk field on the left is found by adding the components shown on the right, cell by cell, into the inverse logit function shown in Equation (\ref{lin}).}
    \label{fig:simPlot}
\end{figure}

\begin{figure}[ht]
    \centering
    \includegraphics[width=1.0\textwidth]{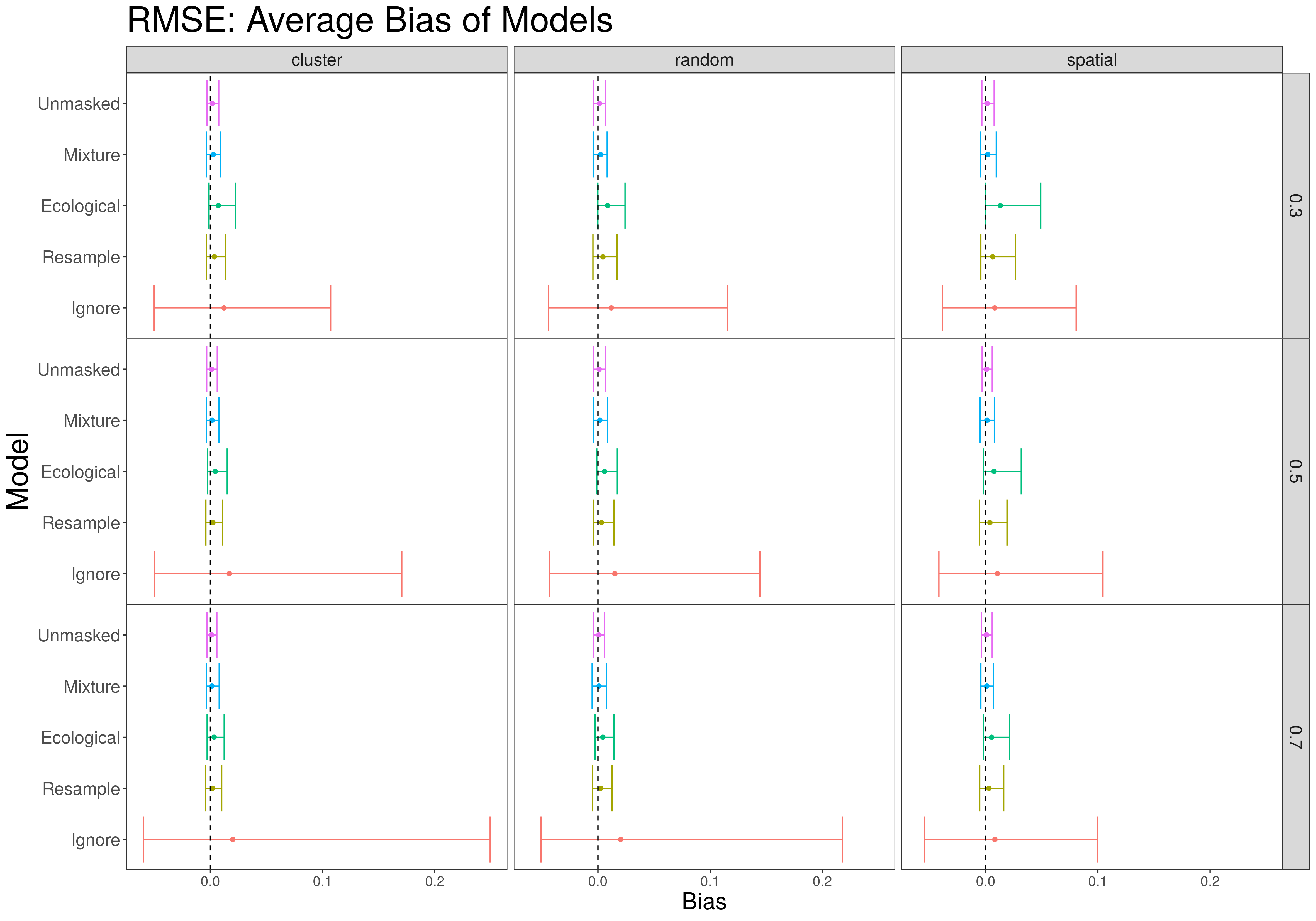}
    \caption{Mean and 95\% quantile range for bias measures for cell level estimates of risk field across all simulations.}
    \label{fig:app2}
\end{figure}

\begin{figure}[ht]
    \centering
    \includegraphics[width=1.0\textwidth]{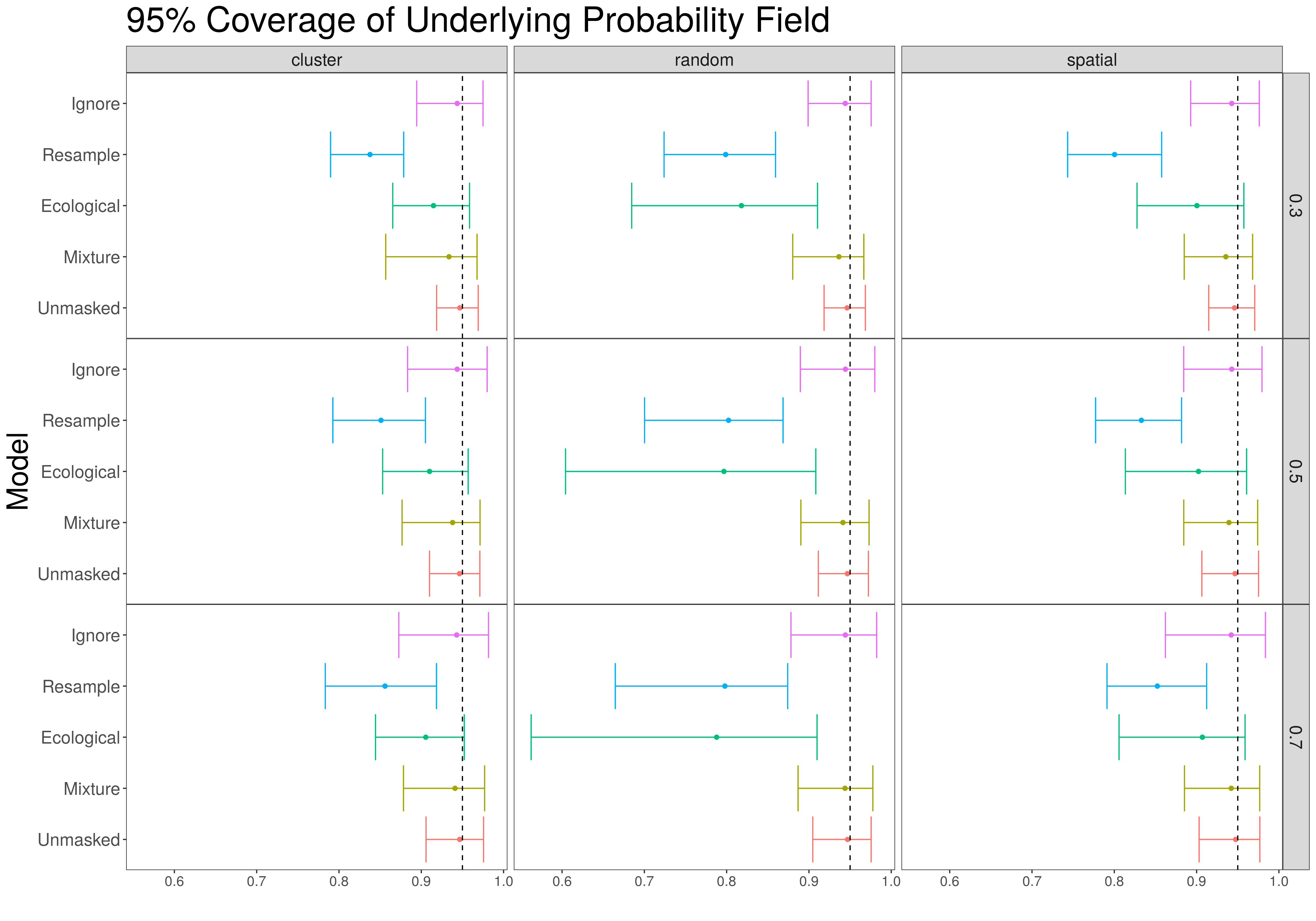}
    \caption{Mean and 95\% quantile range for 95\% confidence/credible intervals for cell level estimates of risk field across all simulations.}
    \label{fig:app1}
\end{figure}

\clearpage

\newpage

\subsection{Appendix D: Dominican Republic Simulation Results}

\begin{figure}[ht]
    \centering
    \includegraphics[width=1.0\textwidth]{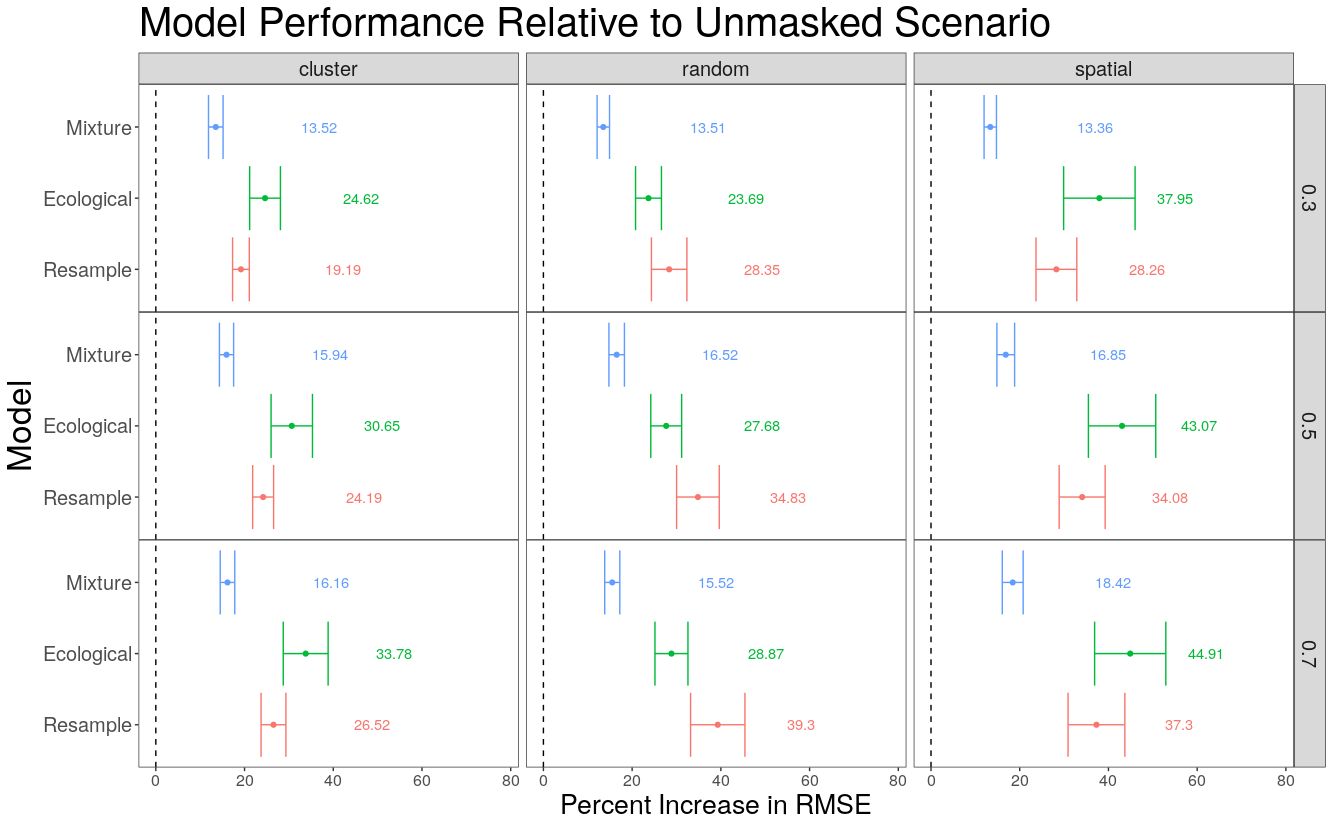}
    \caption{Model performance relative to the gold standard of an unmasked method for Dominican Republic Simulations. Values represent the percent increase of the RMSE compared to the unmasked method. A value of 0\% indicates a method which has the same RMSE as the unmasked method while 100\% indicates a method whose RMSE is twice that of the unmasked method. The vertical bars represent the 2.5 and 97.5 percentiles of results. Ignore model removed from graph due to poor performance and difficulty of comparison to other groups.}
    \label{fig:app3}
\end{figure}

\newpage

\begin{figure}[ht]
    \centering
    \includegraphics[width=1.0\textwidth]{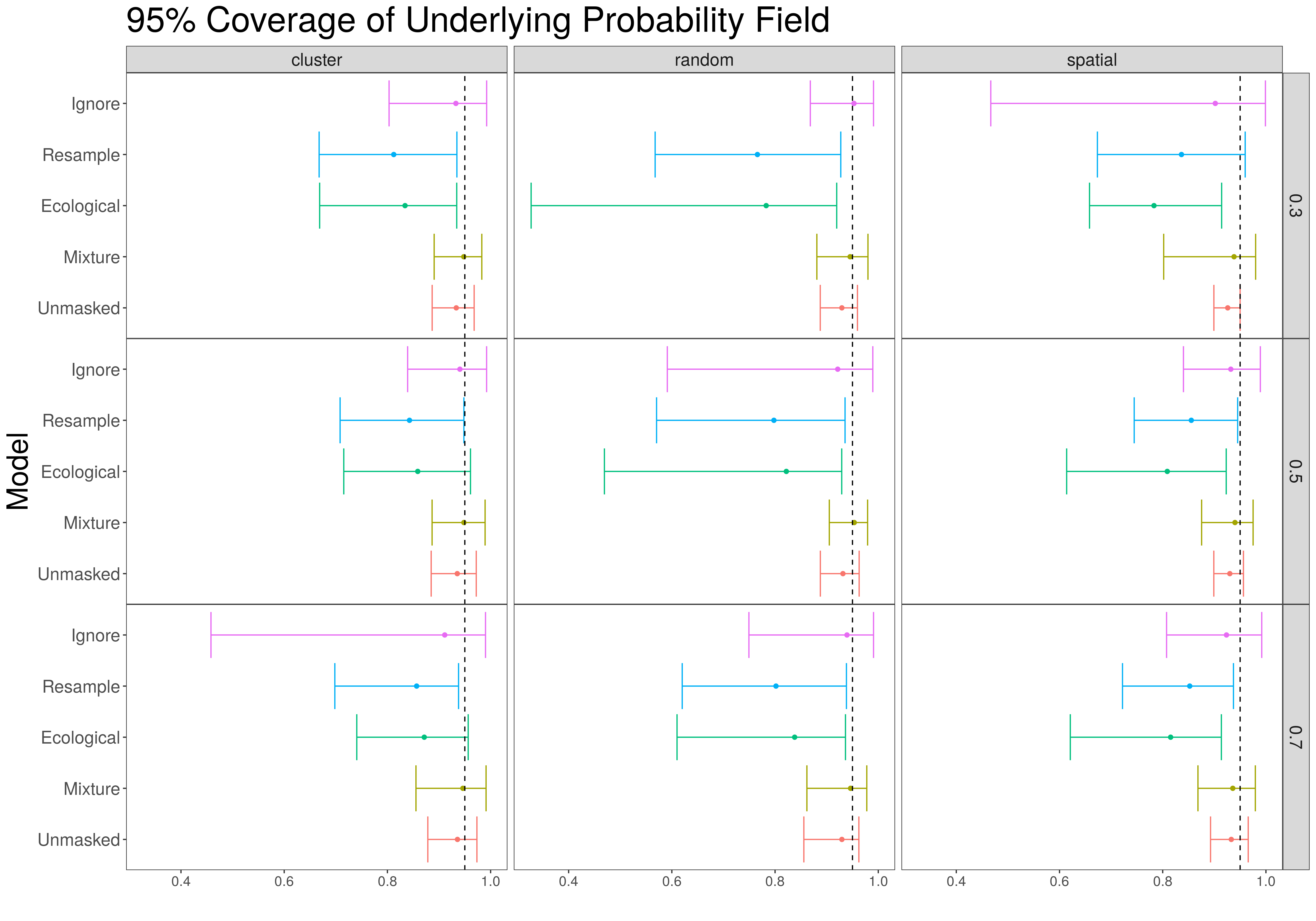}
    \caption{Mean and 95\% quantile range for 95\% confidence/credible intervals for cell level estimates of risk field across all simulations.}
    \label{fig:app4}
\end{figure}

\newpage

\begin{figure}[ht]
    \centering
    \includegraphics[width=1.0\textwidth]{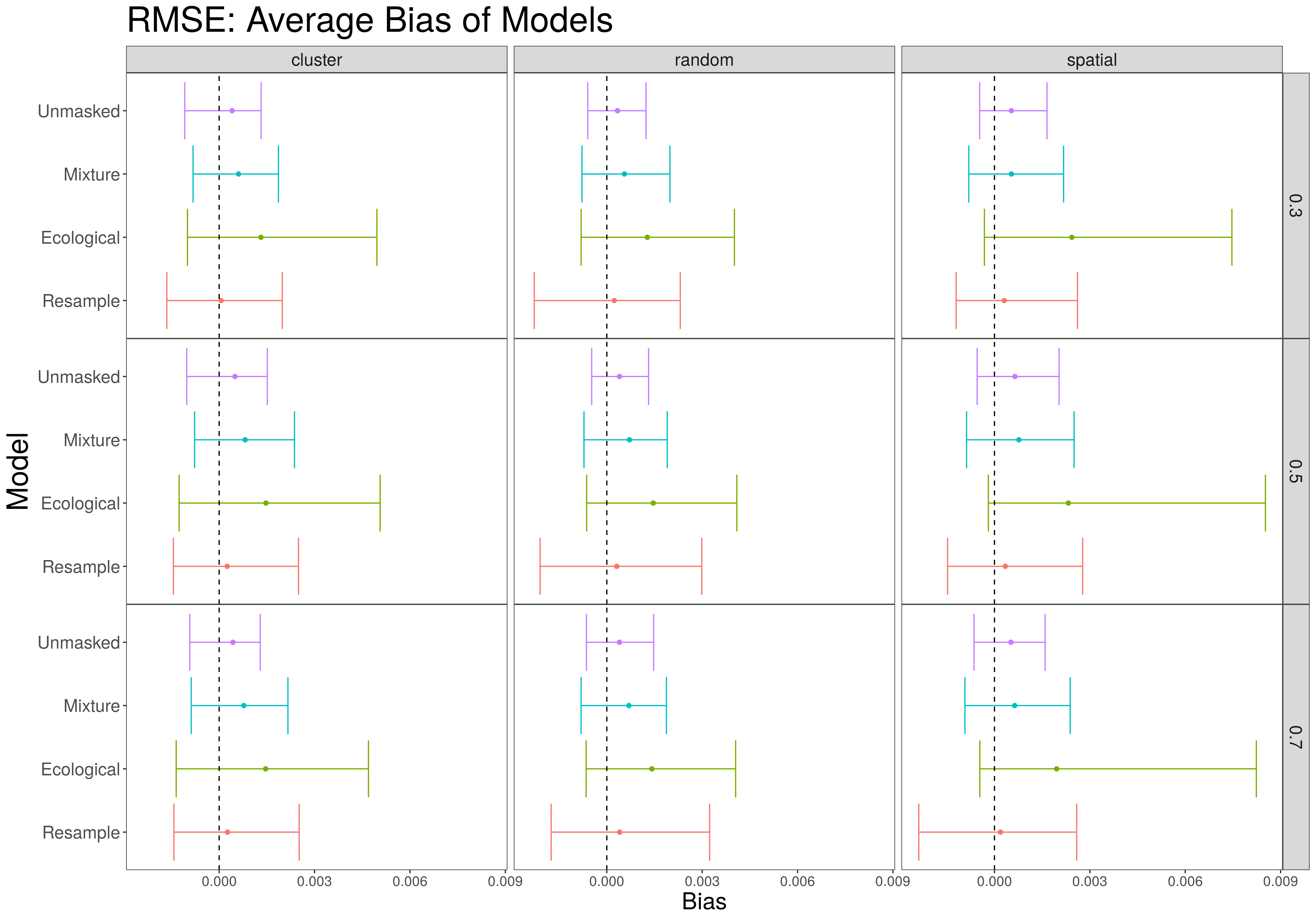}
    \caption{Mean and 95\% quantile range for bias measures for cell level estimates of risk field across all simulations. Ignore model removed from graph due to poor performance and difficulty of comparison to other groups.}
    \label{fig:app5}
\end{figure}

\newpage

\begin{figure}[ht]
    \centering
    \includegraphics[width=1.0\textwidth]{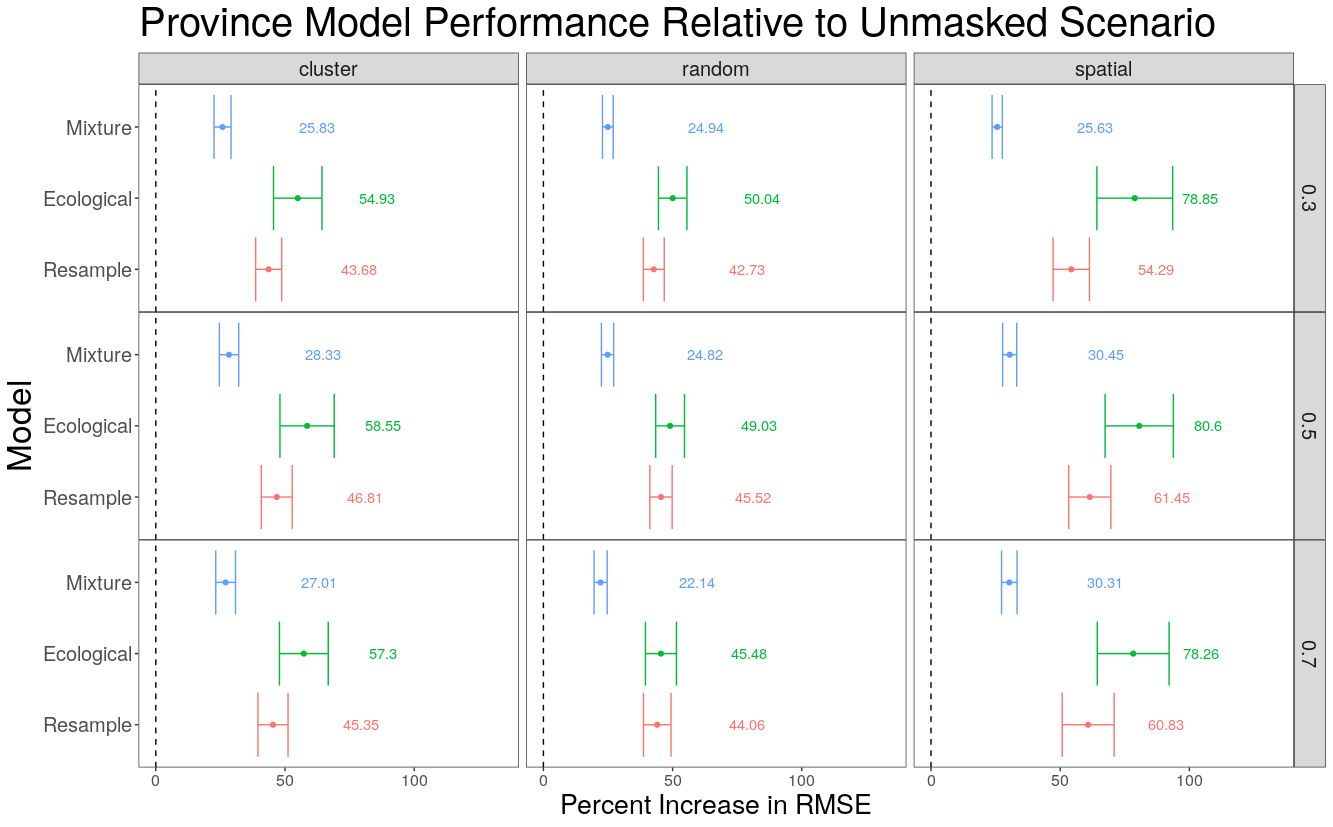}
    \caption{Model performance relative to the gold standard of an unmasked method for Dominican Republic Simulations. Values represent the percent increase of the RMSE compared to the unmasked method for province level risk. A value of 0\% indicates a method which has the same RMSE as the unmasked method while 100\% indicates a method whose RMSE is twice that of the unmasked method. The vertical bars represent the 2.5 and 97.5 percentiles of results. Ignore model removed from graph due to poor performance and difficulty of comparison to other groups.}
    \label{fig:app6}
\end{figure}

\newpage

\begin{figure}[ht]
    \centering
    \includegraphics[width=1.0\textwidth]{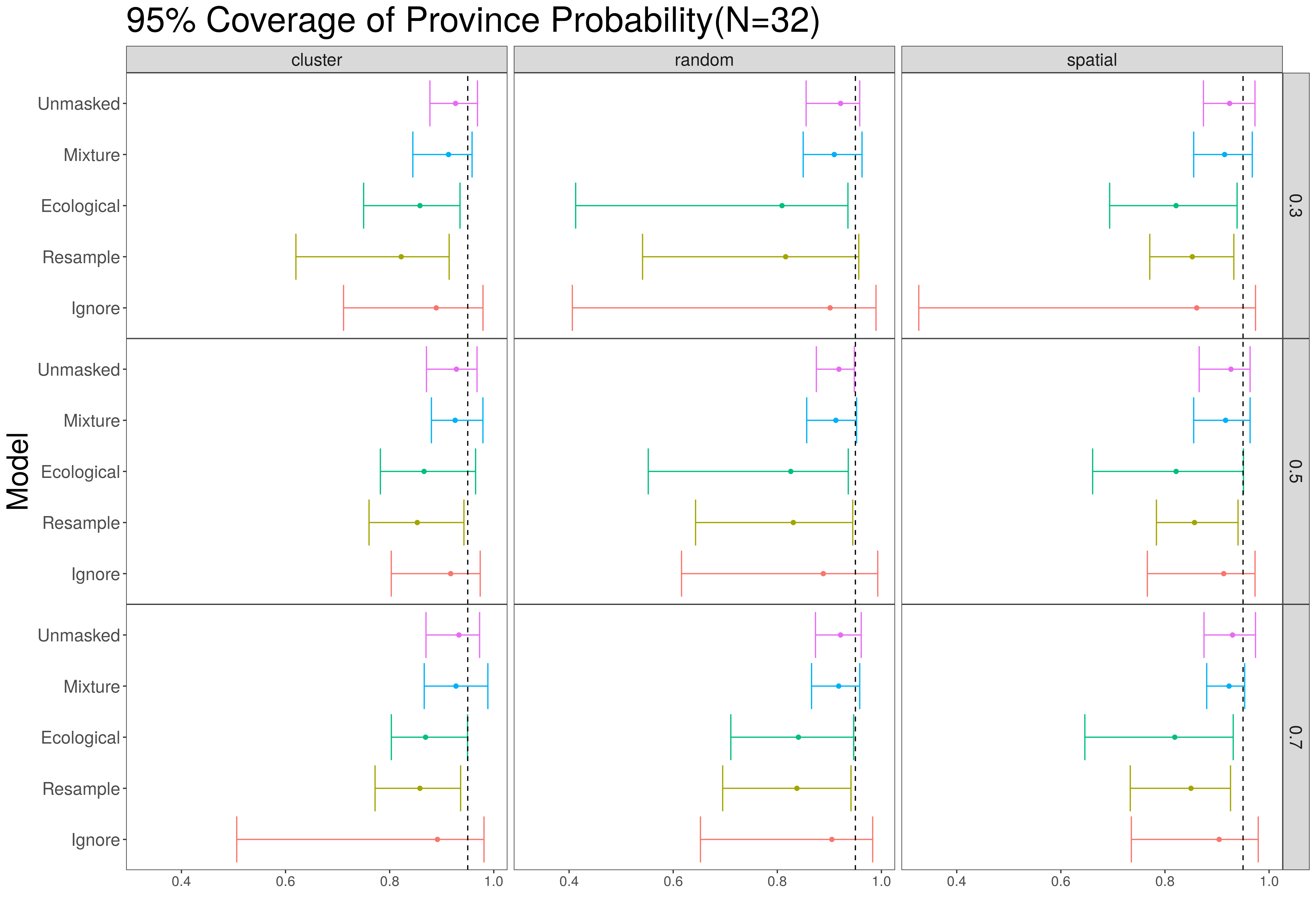}
    \caption{Mean and 95\% quantile range for 95\% confidence/credible intervals for province level estimates of risk field across all simulations, for $N=32$ provinces.}
    \label{fig:app7}
\end{figure}

\clearpage

\subsection{Appendix E: Comparison of U5MR Results for the Dominican Republic}

To compare the 12 approaches we use the conditional predictive ordinate (CPO) and create 14 test datasets by sequentially removing a years data at a time, and comparing the left out data $\boldsymbol{y}_t$  with the predictive distribution based on the data from the other 13 years $\boldsymbol{y}_{-t}$. For approach $a$ and year $t$, we calculate,
$$\mbox{CPO}_t^a = \Pr_a(\boldsymbol{y}_t | \boldsymbol{y}_{-t})$$
for $a=1,\dots,12$ and $t=1,\dots,14$.

Over all years we have the summary:
$$\mbox{CPO}^a = \prod_{t=1}^{14}{\Pr}_a(\boldsymbol{y}_t | \boldsymbol{y}_{-t}).$$
Suppose in year $t$ there are $C_t$ left out (test) clusters. Then
\begin{eqnarray*}
{\Pr}_a(\boldsymbol{y}_t | \boldsymbol{y}_{-t}) &=& {\Pr}_a(y_{t1},\dots,y_{tC_t} | \boldsymbol{y}_{-t})\\
&=&
\int \cdots \int {\Pr}_a(y_{t1},\dots,y_{tC_t} | p^a_{t1},\dots,p^a_{tC_t}) \pi ( p^a_{t1},\dots,p^a_{tC_t} | \boldsymbol{y}_{-t})~dp^a_{t1}\dots dp^a_{tC_t}\\
&=& \prod_{c=1}^{C_t} \int  {\Pr}_a(y_{tc}| p^a_{tc}) \pi ( p^a_{tc} | \boldsymbol{y}_{-t})~dp^a_{tc}
\end{eqnarray*}
where ${\Pr}_a({y}_{tc} | p^{a}_{tc}) $ is a Binomial distribution, and $p^{a}_{tc}$ depends on the approach taken, $a$.
With samples $p^{a}_{tc}$ from the posterior $\pi ( p^a_{tc} | \boldsymbol{y}_{-t})$, $s=1,\dots,S$,  we have the approximation,
\begin{eqnarray*}
\widehat {\Pr}_a(\boldsymbol{y}_t | \boldsymbol{y}_{-t})
&\approx & \prod_{c=1}^{C_t} {\Pr}_a(\boldsymbol{y}_{tc} | p^{a)}_{tc}). 
\end{eqnarray*}
It is usual to report
$$\log (\mbox{CPO}^a)  = \sum_{t=1}^{14}\log {\Pr}_a(\boldsymbol{y}_t | \boldsymbol{y}_{-t}),$$
and these are the numbers we report in Table \ref{tab:cpo}, along with the yearly contributions.

\begin{table}[ht]
\caption{Values of log(CPO) by fitting method. Worst performing models are in {\color{red}{red}} and best in {\color{green}{green}}.}
\centering
\begin{tabular}{rrrrr}
  \hline
 & Year & Mixture & Ignore & Resample \\ 
  \hline
1 & 2000 & -556.78 & \color{red}{-558.10} & \color{green}{-556.42} \\ 
  2 & 2001 & \color{green}{-574.16} & \color{red}{-575.62} & -574.98 \\ 
  3 & 2002 & \color{red}{-547.02} & -546.78 & \color{green}{-546.57} \\ 
  4 & 2003 & -425.10 & \color{red}{-426.13} & \color{green}{-424.94} \\ 
  5 & 2004 & -491.78 & \color{green}{-491.57} & \color{red}{-493.19} \\ 
  6 & 2005 & \color{green}{-532.96} & -533.16 & \color{red}{-533.65} \\ 
  7 & 2006 & -502.69 & \color{red}{-503.49} & \color{green}{-502.20} \\ 
  8 & 2007 & \color{green}{-261.61} & -261.78 & \color{red}{-263.34} \\ 
  9 & 2008 & \color{green}{-106.24} & -106.62 & \color{red}{-107.09} \\ 
  10 & 2009 & -85.34 & \color{red}{-85.58} & \color{green}{-85.21} \\ 
  11 & 2010 & \color{green}{-116.66} & -116.96 & \color{red}{-117.59} \\ 
  12 & 2011 & -120.30 & \color{green}{-120.24} & \color{red}{-121.98} \\ 
  13 & 2012 & \color{green}{-122.30} & -122.54 & \color{red}{-124.34} \\ 
   \hline
   Total& & \color{green}{-4442.96} & -4448.576 & \color{red}{-4451.494} 
\end{tabular}
\label{tab:cpo}
\end{table}

\begin{figure}[ht]
    \centering
    \includegraphics[width=.9\textwidth]{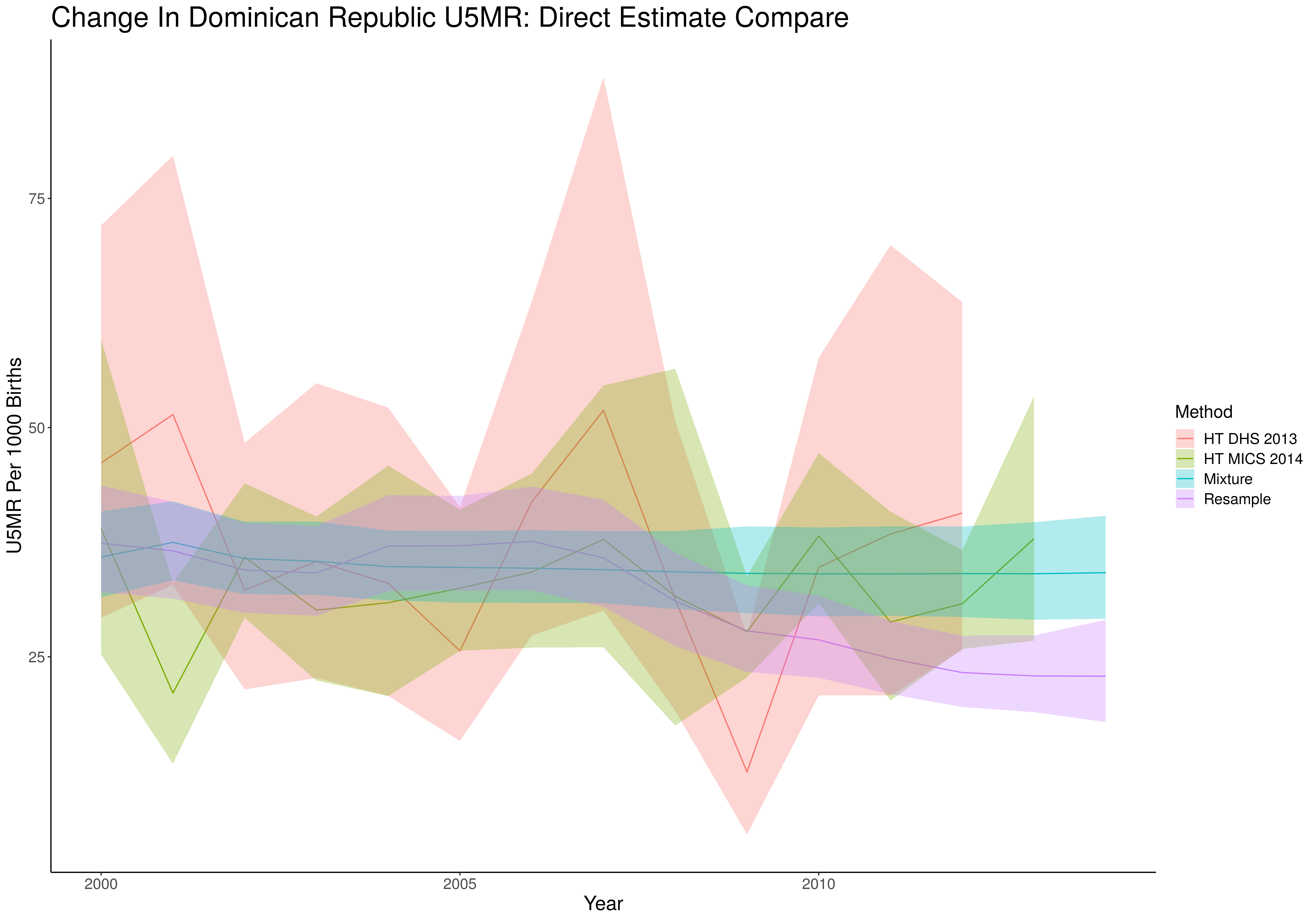}
    \caption{National level estimates of the U5MR over time, from candidate models and direct estimates.}
    \label{fig:app8}
\end{figure}

\begin{figure}[ht]
    \centering
    \includegraphics[width=.9\textwidth]{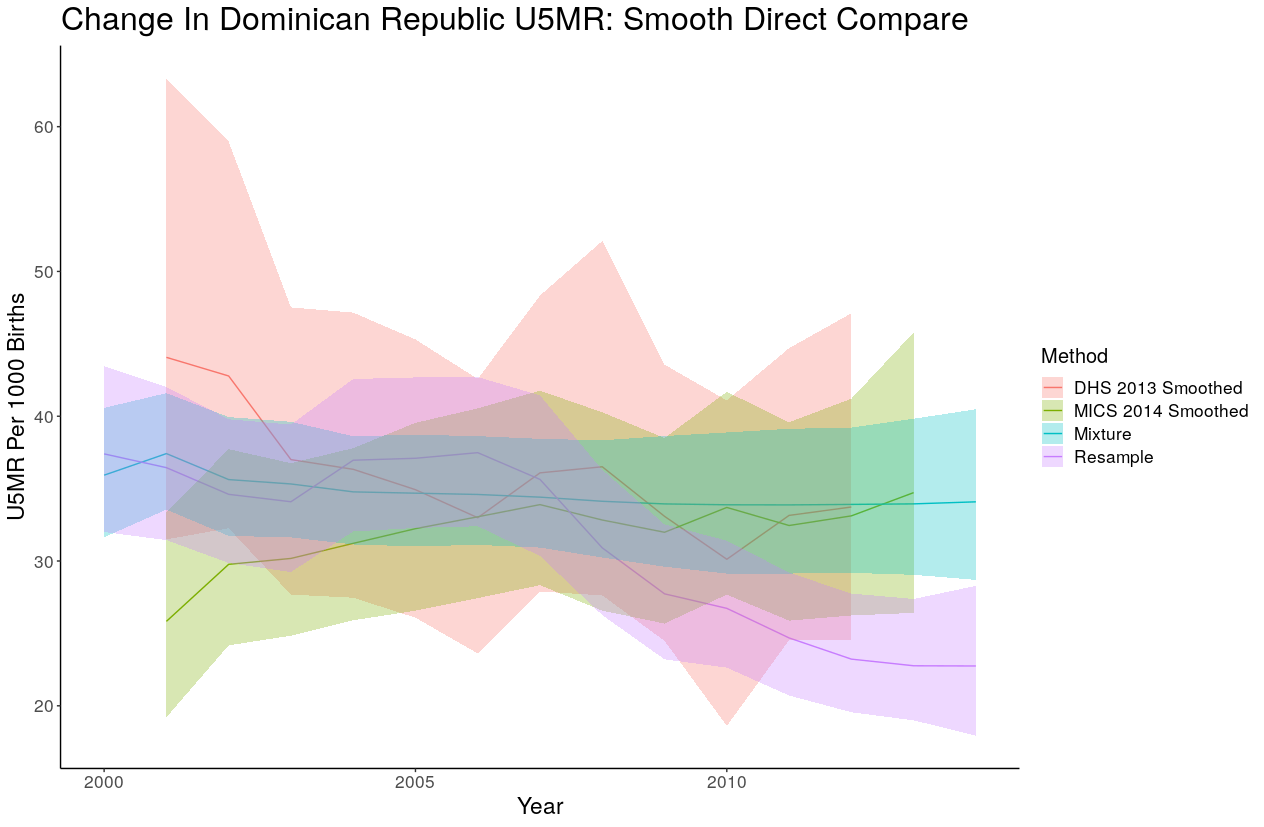}
    \caption{National level estimates of the U5MR over time, from candidate models and smoothed direct estimates.}
    \label{fig:app9}
\end{figure}

\end{document}